\documentclass[twocolumn]{aastex631}

\usepackage{ulem}
\usepackage{xcolor}
\usepackage{graphicx}   
\usepackage{amsmath}    
\usepackage{amssymb}    
\usepackage{diagbox}    
\usepackage{subfigure}
\usepackage{float}
\usepackage{booktabs}
\usepackage{threeparttable}
\usepackage{color}
\usepackage{subfigure}
\usepackage{cancel}
\usepackage{soul}
\usepackage{graphicx}
\usepackage{float}
\usepackage{natbib}
\usepackage{dcolumn}
\usepackage{bm}
\usepackage{ulem}
\usepackage{color}
\usepackage{txfonts}
\usepackage{rotating}
\usepackage{booktabs}
\usepackage{array}
\usepackage{tabularx}
\usepackage{threeparttable}
\usepackage{diagbox}
\usepackage{multirow}
\usepackage{makecell}
\usepackage{longtable}

\shorttitle{A possible formation scenario of the Gaia ID 3425577610762832384: inner binary merger in triple common envelope}
\shortauthors{Zhuowen Li et al.}

\begin{document}

\title{A possible formation scenario of the Gaia ID 3425577610762832384: inner binary merger inside a triple common envelope}

\correspondingauthor{Chunhua Zhu, Guoliang L\"{u}}
\email{chunhuazhu@sina.cn, guolianglv@xao.ac.cn}

\author[0009-0006-1716-357X]{Zhuowen Li}
\email{ZhuoWenli2024@163.com}
\affiliation{School of Physical Science and Technology, Xinjiang University, Urumqi, 830046, China}

\author[0000-0002-3849-8962]{Xizhen Lu}
\affiliation{School of Physical Science and Technology, Xinjiang University, Urumqi, 830046, China}

\author[0000-0002-3839-4864]{Guoliang L\"{u}}
\affil{School of Physical Science and Technology,
Xinjiang University, Urumqi, 830046, China}
\affil{Xinjiang Observatory,
the Chinese Academy of Sciences, Urumqi, 830011, China}

\author{Chunhua Zhu}
\affiliation{School of Physical Science and Technology, Xinjiang University, Urumqi, 830046, China}

\author{Helei Liu}
\affiliation{School of Physical Science and Technology, Xinjiang University, Urumqi, 830046, China}



\author[0000-0001-8493-5206]{Jinlong Yu}
\affiliation{College of Mechanical and Electronic Engineering, Tarim University, Alar, 843300, China}

\begin{abstract}
Recently, an identified non-interacting black hole (BH) binary, Gaia ID 3425577610762832384 (hereafter G3425), contains a BH ($\sim$3.6 M$_{\odot}$) falling within the mass gap and has a nearly circular orbit, challenging the classical binary evolution and supernova theory. Here, we propose that G3425 originates from a triple through a triple common envelope (TCE) evolution. The G3425 progenitor originally may consist of three stars with masses of 1.49 M$_{\odot}$, 1.05 M$_{\odot}$, and 21.81 M$_{\odot}$, and inner and outer orbital periods of 4.22 days and 1961.78 days, respectively. As evolution proceeds, the tertiary fills its Roche lobe, leading to a TCE. We find that the orbital energy generated by the inspiral of the inner binary serves as an additional energy imparted for ejecting the common envelope (CE), accounting for $\sim$97\% of the binding energy in our calculations. This means that the outer orbit needs to expend only a small amount of the orbital energy to successfully eject CE. The outcome of the TCE is a binary consisting of a 2.54 M$_\odot$ merger produced by the inner binary merger and a 7.67 M$_\odot$ helium star whose CE successfully ejected, with an orbital period of 547.53 days. The resulting post-TCE binary (PTB) has an orbital period that is 1-2 orders of magnitude greater than the orbital period of a successfully ejected classical binary CE. In subsequent simulations, we find that the successfully ejected helium star has a 44.2\% probability of forming a BH. In the case of a non-complete fallback forming a BH, with an ejected mass of 2.6 M$_{\odot}$ and a relatively low natal kick ($11^{+16}_{-5}$ ${\rm km/s}$ to $49^{+39}_{-39}$ ${\rm km/s}$), this PTB can form G3425 in the Milky Way.
\end{abstract}


\section{Introduction}
Black hole (BH) binaries are excellent laboratories for understanding massive star evolution, binary evolution, and Supernovae (SNe). So far, most observed BH binaries are BH X-ray binaries \citep{2006csxs.book..157M,2006ARA&A..44...49R,2014SSRv..183..223C,2016A&A...587A..61C}, which are also the most studied type of BH binaries (e.g., \cite{2009ApJ...707..870B}, \cite{2015ApJ...809...99S}, \cite{2018MNRAS.481.1908K}, \cite{2018MNRAS.479.4391M}, and \cite{2020ApJ...898..143S}). However, based on the observed outburst characteristics and distance distribution of known BH X-ray binaries, their number is expected to represent only a small fraction of the entire BH binary population \citep{2016A&A...587A..61C}.

With the rapid advancement of astrometric instruments and technology, non-interacting BH binaries are also gradually being unveiled. Very recently, Gaia DR3, the pre-release of Gaia DR4, LAMOST, and Gaia DR2 have confirmed four non-interacting BH binaries using spectroscopic and astrometric data, named Gaia BH1 \citep{2023AJ....166....6C,2023MNRAS.518.1057E}, Gaia BH2 \citep{2023MNRAS.521.4323E,2023ApJ...946...79T}, Gaia BH3 \citep{2024A&A...686L...2G}, and Gaia ID 3425577610762832384 (hereafter G3425) \citep{2024NatAs.tmp..215W}. G3425 has many unique features, and some of its physical properties are listed in Table \ref{tab:1}. Compared to known BH X-ray binaries, G3425 has a much longer orbital period. Additionally, the BH in G3425 has a much lower mass compared to the BHs in Gaia BH1, Gaia BH2, and Gaia BH3, falling within the 3$\sim$5 M$_{\odot}$ range. BHs within this mass range are rare in known BH binaries, the range often referred to as the mass gap \citep{1997AAS...190.1001B,2001ApJ...554..548F,2010ApJ...725.1918O,2011ApJ...741..103F}. Furthermore, G3425 has a lower eccentricity compared to Gaia BH1, Gaia BH2, and Gaia BH3, with its orbit being closer to circular.

G3425 challenges the classical binary evolution theory. In the discussion of the isolated binary origin of G3425 by \cite{2024NatAs.tmp..215W}, they consider the high mass ratio between the BH progenitor and the visible giant in G3425. If the progenitor binary underwent mass transfer (MT), it is likely that it went through the common envelope evolution (CEE) phase. In the CEE simulations by \cite{2024NatAs.tmp..215W}, forming G3425 typically requires an excessively large ejection efficiency parameter ($\alpha_{\rm CE}$), with typical $\alpha_{\rm CE}$ values ranging from 5 $\sim$ 10. Recently, \cite{2024MNRAS.535L..44G} and \cite{2024arXiv241018501K} proposed that increasing the overshooting parameter or stellar wind strength in massive stars could suppress the expansion of their radii, potentially allowing the progenitor binary of G3425 to avoid undergoing Roche-lobe overflow (RLOF). However, in the simulations by \cite{2024NatAs.tmp..215W}, even with a tenfold increase in stellar wind strength, the progenitor of this BH still fills its Roche lobe radius. On the other hand, in the analysis by \cite{2024NatAs.tmp..215W}, G3425 is suggested to possibly originate from a triple, where the BH is formed from the merger of two neutron stars (NS) or a Thorne-$\dot{\rm Z}$ytkow object (the product of the merger of a NS with a giant star \citep{1995MNRAS.274..485P}). However, in the triple population synthesis analysis by \cite{2022MNRAS.516.1406S}, no surviving triples were found with an inner binary consisting of binary NS, unless it is assumed that zero natal kicks during the formation of NS.

It is well known that a high fraction of massive stars are born in triple or multiple stars \citep{2013A&A...550A.107S,2017ApJS..230...15M}. Based on the observational statistics of \cite{2017ApJS..230...15M}, more than $\sim$70\% of massive stars are in triple or higher-order (e.g., quadruple) configurations. Hierarchical triples are known for their long-term ZLK oscillations, which are caused by the exchange of angular momentum between the inner and outer orbits \citep{1910AN....183..345V,1962AJ.....67..591K,1962P&SS....9..719L,2016ARA&A..54..441N}. This leads to the excitation of the eccentricity and inclination of the inner orbit, which ultimately enhances tidal effects, gravitational wave emission, and inner binary interactions (e.g., mass transfer and collisions) \citep{2014ApJ...793..137N,2016ComAC...3....6T,2016ApJ...822L..24N,2019MNRAS.487.3029S,2020A&A...640A..16T,2023ApJ...955L..14S,2024arXiv240811128B}. Recent studies also suggest that the ZLK mechanism in triples can explain the origin of events such as low-mass X-ray binaries \citep{2016ApJ...822L..24N,2024arXiv241115644S}, high-velocity runaway stars \citep{2019MNRAS.487.3029S,2024arXiv240811128B}, type Ia supernovae \citep{2018A&A...610A..22T,2023ApJ...955L..14S,2023ApJ...950....9R}, blue straggler binaries \citep{2014ApJ...793..137N,2024arXiv240706257S}, NS and white dwarf merger \citep{2018A&A...619A..53T}, and binary BH mergers \citep{2017ApJ...841...77A,2020ApJ...903...67M}. In addition, the interaction of triple RLOF in hierarchical triples has gained increasing attention \citep{2014MNRAS.438.1909D,2020MNRAS.498.2957C,2021MNRAS.500.1921G,2022ApJS..259...25H,2023ApJ...950....9R,2024arXiv240903826K,2024Natur.635..316B}. During this phase, the tertiary overflows its Roche lobe, transferring mass to the inner binary. Triple RLOF can lead to either stable or unstable MT. In the case of unstable MT, it can result in a triple common envelope (TCE), where the extended envelope engulfs the inner binary and the core of the tertiary. During the TCE process, the inner binary inspiral each other and towards the core of the donor due to friction \citep{2013A&ARv..21...59I,2020cee..book.....I,2023LRCA....9....2R}. TCE can result in various possible outcomes, such as the merger of the inner binary, the ejection of one star (usually the least massive component), and chaotic triple dynamics, among others \citep{2015MNRAS.450.1716S,2020MNRAS.498.2957C,2021MNRAS.504.5967S,2021MNRAS.500.1921G,2022ApJS..259...25H,2024arXiv240811128B}. Here, we propose that G3425 may originate from a hierarchical triple, which underwent TCE. In this scenario, the tertiary is the progenitor of the BH, while the giant evolved from the merger product of the inner binary. If the contribution of the orbital energy of the inner binary to CE ejection is considered during the TCE process, the outer orbit may not require excessive energy to successfully eject CE. In other words, the outer orbit does not need to spiral in as deeply as in a binary CEE. Typically, in a binary CEE process, for $\alpha_{\rm CE}$ = 1, the post-CEE orbital period is $<$ $\sim$ 10 days \citep{2013A&ARv..21...59I,2020cee..book.....I,2023LRCA....9....2R}. The results suggest that post-TCE binary (PTB) may have longer orbital period, which potentially could explain G3425.

The structure of this $Letter$ is as follows. In Section 2, we describe the modeling of the evolution of the progenitor triple of G3425, the TCE process, the evolution of the PTB, and the modeling of SNe. In Section 3, we present the computational results for the formation of G3425, followed by a conclusion in Section 4.

\begin{table}[]
    \caption{Physical parameters of the observed G3425. The second and third rows provide the masses of the optical companion and BH, respectively. The optical companion is a red giant (RG) star. Rows four and five show the orbital period and eccentricity, respectively. The sixth row and the last row represent the metallicity and the ratio of the RG's radius to its Roche lobe radius, respectively. The data come from \cite{2024NatAs.tmp..215W}.}
    \begin{tabular}{ll}
    \hline
        Physical parameter & G3425 \\ \hline
        $M_{\rm RG}$ (M$_{\odot}$) & $2.66^{+1.18}_{-0.68}$ \\
        $M_{\rm BH}$ (M$_{\odot}$) & $3.6^{+0.8}_{-0.5}$ \\
        $P_{\rm orb}$ (days) & $877^{+2}_{-2}$ \\
        $e$ & $0.05^{+0.01}_{-0.01}$ \\
        $[{\rm Fe}/{\rm H}]$ & $-0.12^{+0.02}_{-0.02}$ \\
        $R_{\rm RG}$/$R_{\rm RG,L}$ & $\sim$4.5\% \\
        \hline
    \end{tabular}
    \label{tab:1}
\end{table}

\section{Methodology}
The formation of G3425 in our simulation is divided into several sub-processes. It starts with the evolution of the initial triple until the TCE occurs, then the modeling of the TCE process and the evolution of the PTB, and finally the occurrence of SNe and the evolution of the post-SNe binary (up to Hubble time). In the following subsections, we explain the modeling methods of these sub-processes in detail.

\subsection{Modeling of triple evolution and triple common envelope}
Following \cite{2021MNRAS.502.4479H}, \cite{2022ApJ...925..178H} and \cite{2024ApJ...975L...8L}, we require the initial triple to be dynamically stable. Specifically, we require the initial triple to satisfy the formula from \cite{2001MNRAS.321..398M}, which is:
\begin{equation}
\frac{a_{\rm out}}{a_{\rm in}}>2.8\left(1+\frac{m_{3}}{m_{1}+m_{2}}\right)^{\frac{2}{5}} \frac{\left(1+e_{\rm out}\right)^{\frac{2}{5}}}{\left(1-e_{\rm out}\right)^{\frac{6}{5}}}\left(1-\frac{0.3 i}{180^{\circ}}\right)
\end{equation}
Here, for a hierarchical triple, two stars are on a tighter orbit (the inner binary), while the third companion orbits the inner binary on a wider orbit (i.e., the center of mass of the inner binary and the tertiary form an outer binary). The subscripts "in" and "out" denote the inner and outer parts of the triple, respectively. Subscripts 1, 2, and 3 refer to the primary, the secondary in the inner binary, and third component, respectively. The symbol $i$ represents the mutual inclination between the pairs of orbits. Additionally, we reject initial inner binaries that are in RLOF at periastron, by using \cite{1983ApJ...268..368E} analytical formula to calculate the Roche lobe radius and $R_{\rm ini} = R_{\odot}(m_{\rm ini}/M_{\odot})^{0.7}$ \citep{1994sse..book.....K} to estimate the initial stellar radius. For the initial triples that meet the above criteria, we use the Multiple Stellar Evolution (MSE) code \citep{2021MNRAS.502.4479H} to simulate their evolution.
The advantage of the MSE code is that it includes rapid fitting formulas for single star evolution \citep{2000MNRAS.315..543H}, binary interactions (e.g., tidal effects, mass transfer) \citep{2002MNRAS.329..897H}, fly-bys and dynamical perturbations in multiple systems. For the long-term dynamical evolution of multiple systems, the MSE code uses orbital-averaged integration when the system is sufficiently hierarchical \citep{2016MNRAS.459.2827H,2018MNRAS.476.4139H,2020MNRAS.494.5492H}, and self-consistent modeling through N-body methods when the system is dynamically unstable \citep{2020MNRAS.492.4131R}. However, the MSE code still has certain limitations. For example, the rapid fitting formulas for single star evolution and binary interactions included in the MSE code are extreme approximations. In some cases, they may overshoot the radius evolution of single stars and make approximations during mass transfer evolution. Additionally, these rapid fitting formulas do not model stellar structure. Considering the high metallicity of the visible companion of G3425, the triple is set with a typical Galactic metallicity of Z = 0.014 \citep{2012A&A...537A.146E}. Moreover, for parameter values in the MSE code that are not mentioned in this paper, we use the default values.

When the tertiary fills its Roche lobe and MT occurs, we follow the default settings of the MSE code, using the critical mass ratio to determine whether the MT is stable. If the MT is unstable, it will lead to TCE. However, the evolution of TCE is still highly uncertain, as it often requires detailed modeling involving higher dimensions, such as hydrodynamics and gravitational dynamics \citep{2015MNRAS.450.1716S,2020MNRAS.498.2957C,2021MNRAS.504.5967S,2021MNRAS.500.1921G,2022ApJS..259...25H}. In one of the three-dimensional numerical simulation results by \cite{2021MNRAS.500.1921G} on TCE, both the inner and outer orbits undergo inspiral due to friction, and their orbital energy changes ($\Delta E_{\rm orb}$) contribute to the ejection of the CE. Therefore, we use the standard energy prescription to first assume that both the inner and outer orbital energies contribute to the consumption of the binding energy ($E_{\rm bind}$). The specific formula is as follows \citep{1976IAUS...73...75P,1976IAUS...73...35V,1984ApJ...277..355W,1988ApJ...329..764L,1993PASP..105.1373I,2013A&ARv..21...59I}:
\begin{equation}
E_{\rm bind}=\alpha_{\rm CE, in} \Delta E_{\rm orb, in}+\alpha_{\rm CE, out} \Delta E_{\rm orb, out}
\end{equation}
Here, $E_{\rm bind}$ of the envelope is contributed by the envelope of the tertiary. The changes in inner orbital energy $\Delta E_{\rm orb,in}$ and outer orbital energy $\Delta E_{\rm orb,out}$ are calculated as the differences between the orbital energies before and after the inspiral, respectively, as follows:
\begin{equation}
\left\{\begin{array}{c}
E_{\rm bind}=G \frac{m_{\rm 3} m_{\rm 3, env}}{\lambda R_{\rm 3}} \\
\Delta E_{\rm orb, in}=-\frac{G m_{\rm 1} m_{\rm 2}}{2 a_{\rm in, i}}+\frac{G m_{\rm 1} m_{\rm 2}}{2 a_{\rm in, f}} \\
\Delta E_{\rm orb, out}=-\frac{G m_{\rm 3}\left(m_{\rm 1}+m_{\rm 2}\right)}{2 a_{\rm out, i}}+\frac{G m_{\rm 3, core}\left(m_{\rm 1}+m_{\rm 2}\right)}{2 a_{\rm out, f}}
\end{array}\right.
\end{equation}
Here, ${\rm G}$ represents the gravitational constant. Subscripts ${\rm i}$ and ${\rm f}$ denote the states before and after the inspiral, respectively, while subscripts ${\rm env}$ and ${\rm core}$ refer to the envelope (hydrogen-rich envelope) and core (helium core) of the donor star (here, the tertiary). The parameters $\alpha_{\rm CE,in}$ and $\alpha_{\rm CE,out}$ represent the ejection efficiency for the inner and outer orbital energies. Based on the general settings used in many previous population synthesis calculations (e.g., \cite{2018MNRAS.474.2959G}, \cite{2020ApJ...898..143S}, \cite{2021ApJ...920...81S}, \cite{2022ApJ...931..107C}, and \cite{2024ApJ...975..163C}), we set $\alpha_{\rm CE,in}=\alpha_{\rm CE,out}=1$. Although $\alpha_{\rm CE}$ is very important in TCE evolution, with larger values making it easier to eject CE during the TCE process \citep{2021MNRAS.500.1921G}, it remains highly uncertain \citep{2010A&A...520A..86Z,2013A&ARv..21...59I,2023LRCA....9....2R}. Recent numerical simulations of CEE suggest that $\alpha_{\rm CE}$ may range from 0.5 to 2.12 \citep{2024A&A...691A.244V}. $\lambda$ represents the structural parameters of the envelope, for which we use a typical value of $\sim$0.1 for massive stars \citep{2010ApJ...716..114X,2011ApJ...743...49L,2018MNRAS.474.2959G}. Additionally, in one of the simulation results from the study by \cite{2021MNRAS.500.1921G}, the inner binary inspirals faster than the outer binary. Therefore, we also assume that during the TCE process, $a_{\rm in}$ decreases faster than $a_{\rm out}$ ($a_{\rm in}/a_{\rm out}$ keeps decreasing and satisfying Equation 1). This ensures that the triple remains dynamically stable during TCE and that $E_{\rm bind}$ is always consumed first by the inner orbital energy ($E_{\rm orb,in}$), followed by the outer orbital energy ($E_{\rm orb,out}$). Under the above assumptions, using the standard energy prescription, we estimate the following possible outcomes of TCE \citep{2013A&ARv..21...59I,2021MNRAS.500.1921G,2023LRCA....9....2R}:

(i) If $E_{\rm bind} < \Delta E_{\rm orb,in}$, during the TCE process, the CE is ejected, and the inner binary combines with the helium star (the core of the tertiary) forms a new triple.

(ii) If $\Delta E_{\rm orb,in} < E_{\rm bind} < \Delta E_{\rm orb,in} + \Delta E_{\rm orb,out}$, during the TCE process, the CE is ejected, but the inner binary merges, and the merger product of the inner binary combines with the helium star (the core of the tertiary) forms a binary.

(iii) If $E_{\rm bind} > \Delta E_{\rm orb,in} + \Delta E_{\rm orb,out}$, during the TCE process, the CE is not completely ejected. The inner binary merges, and its merger product also merges with the helium star (the core of the tertiary). As a result, the TCE leads to a single star.

We adopt the assumption of \cite{2002MNRAS.329..897H}, \cite{2017PASA...34...58E}, \cite{2018MNRAS.474.2959G}, \cite{2021MNRAS.502.4479H}, \cite{2022ApJS..258...34R}, \cite{2024ApJ...969..160L}, and \cite{2024ApJ...975L...8L} that when $a_{\rm in,f} = R_{1}+R_{2}$, the two stars merge. Additionally, following \cite{1997MNRAS.291..732T} and \cite{2002MNRAS.329..897H}, we assume when two main-sequence (MS) stars merge, their material is fully mixed, and the merger product remains an MS star. We also assume no mass loss during the merger process, meaning the mass of the merger product is $M_{\rm mer}=m_{1}+m_{2}$ \citep{1997MNRAS.291..732T,2002MNRAS.329..897H}. We emphasize that this approach remains highly simplified, but it does capture some of the key results from the TCE simulations by \cite{2021MNRAS.500.1921G}.

\subsection{Modeling of post-triple common envelope and Supernovae}
Using the methods described in the previous section, we calculate the possible outcomes of TCE. In scenarios where the inner binary merges and the outer binary successfully ejects CE, the merger product of the inner binary and the core of the tertiary (a helium star) form the PTB. For the PTB, we use MESA stellar evolution code \citep{2011ApJS..192....3P,2013ApJS..208....4P,2015ApJS..220...15P,2018ApJS..234...34P,2019ApJS..243...10P} (version 10398) to track its evolution. Similar to \cite{2023A&A...674A.216L} and \cite{2023A&A...671A..62Q}, we use the MESA code to create helium-rich stars and then relax the created helium star until the ratio of its helium-burning luminosity to total luminosity exceeds 99\%. We apply the Ledoux criterion and the standard mixing length theory ($\alpha_{\rm MLT} = 1.5$) for convection calculations \citep{1991A&A...252..669L}. The overshooting parameter is set to 0.335 \citep{2011A&A...530A.115B}, and the semi-convection parameter is 1 \citep{1991A&A...252..669L}. For helium-rich stars, its stellar wind scheme is adopted by \cite{2000A&A...360..227N}.

Following \cite{2022A&A...667A..58P}, \cite{2023A&A...674A.216L}, and \cite{2023ApJS..264...45F}, when the helium star evolves to the point of core carbon depletion, we assume it undergoes a SN. To date, SNe still involve significant uncertainties \citep{2012ApJ...749...91F,2016MNRAS.460..742M,2020MNRAS.499.3214M,2021A&A...645A...5S}. For the type and mass of the remnant after a SN, we use the semi-analytical model of \cite{2020MNRAS.499.3214M} for simulation. Specifically, we first determine the remnant probability distribution based on the carbon-oxygen core mass ($M_{\rm CO}$), and the detailed analytical formula is as follows: \citep{2020MNRAS.499.3214M}
\begin{equation}
\left\{\begin{array}{c}
P_{\rm NS} = 1,\ \text{if}\ M_{\rm CO} < 2 M_{\odot} \\
P_{\rm BH} = \frac{M_{\rm CO}-2}{5}\ \text{and}\ P_{\rm NS}=1-P_{\rm BH},\ \text{if}\ 2 M_{\odot} \leq M_{\rm CO} < 7 M_{\odot} \\
P_{\rm BH} = 1,\ \text{if}\ M_{\rm CO} \geq 7 M_{\odot}
\end{array}\right.
\end{equation}
Secondly, if the remnant is a BH, we calculate the probability of it forming through complete fallback (CF), and the specific formula is as follows: \citep{2020MNRAS.499.3214M}
\begin{equation}
\left\{\begin{array}{c}
P_{\rm CF} = 1,\ \text{if}\ M_{\rm CO} \geq 8 M_{\odot} \\
P_{\rm CF} = \frac{M_{\rm CO}-2}{6},\ \text{if}\ 2 M_{\odot} \leq M_{\rm CO} \leq 8 M_{\odot}
\end{array}\right.
\end{equation}
If BH is formed through CF, its mass equals the mass of the helium core (helium stars). Otherwise, its mass is drawn from a normal distribution with a mean of $0.8M_{\rm CO}$ and a standard deviation of 0.5 \citep{2020MNRAS.499.3214M}. We also set the minimum BH mass to 2 M$_{\odot}$. If the generated BH mass falls below this limit, we redraw it until it falls within the allowed range. Nevertheless, we emphasize that our method has certain uncertainties. In the catalogue of BH transients of \cite{2016A&A...587A..61C}, the BH mass function remains uncertain. Furthermore, in the study by \cite{2016ApJ...821...38S}, if the pre-SN star undergoes only helium core collapse, the average BH mass is in the range of 7.7 M$_{\odot}$$\sim$9.2 M$_{\odot}$, and it is predicted that BH cannot form in the mass gap (3 M$_{\odot}$$\sim$5 M$_{\odot}$). On the other hand, observational evidence for BH natal kicks remains quite limited \citep{2024Natur.635..316B}. Many previous studies suggest that the natal kicks received by BH can be large (tens to hundreds of ${\rm km/s}$) \citep{2022MNRAS.517.3938C,2022ApJ...930..159A,2023ApJ...957...68B,2023ApJ...952L..34K,2024arXiv240813310M}, but they can also be very small ($<10\ {\rm km/s}$) \citep{2020PhRvD.101l3013W,2024Ap&SS.369...80J,2024arXiv241016276R,2024arXiv241014778D,2024arXiv241116847N,2024PhRvL.132s1403V} or even zero \citep{2003Sci...300.1119M,2022NatAs...6.1085S}. In previous research, low natal kicks ($<40\sim50\ {\rm km/s}$) are preferred for wide-orbit BH binaries like Gaia BH1 and Gaia BH2 \citep{2023MNRAS.518.1057E,2023MNRAS.521.4323E,2024arXiv240313579K,2024ApJ...975L...8L}. Additionally, very recently, in some observed triples containing a BH (e.g., V404 Cygni), the natal kicks of this BH at formation was almost negligible ($<5\ {\rm km/s}$) \citep{2024Natur.635..316B,2024arXiv241115644S}. Therefore, we assume natal kicks are drawn from Maxwellian distributions with dispersions ($\sigma_{\rm k}$) of 10 ${\rm km/s}$ and 50 ${\rm km/s}$, respectively. However, in some previous studies, the distribution of kicks is sometimes considered to depend on the mass of the BH \citep{2012ApJ...749...91F}. This means that low-mass BH would receive significant kicks, making it harder for the binary to survive. Therefore, using only Maxwellian distributions with $\sigma_{\rm k}$ of 10 ${\rm km/s}$ and 50 ${\rm km/s}$ to describe the kick distribution in this paper may underestimate the kick velocity for low-mass BH. For these two scenarios, we draw $5\times10^{6}$ repetitions of the type of the remnant, the mass of the remnant, orientation angle of SNe, mean anomaly of SNe, and natal kick as a way to investigate whether PTBs can form G3425 during the SNe process. Finally, we use the binary star evolution (BSE) code \citep{2002MNRAS.329..897H} to track the evolution of the post-SNe binary up to Hubble time.

\section{Results}
Studying the formation of G3425 through TCE is crucial for understanding the evolution of massive stars, the evolution of triples, the CEE, and the SNe. In the following subsection, we present the detailed computational results for the formation of G3425.

\subsection{Formation of G3425}
Following \cite{2024ApJ...975L...8L}, we use Monte Carlo simulations to generate the initial input parameters of triples. In the Monte Carlo simulations, we mainly considere both the results of the 3D simulations of TCE by \cite{2021MNRAS.500.1921G} and the observed characteristics of G3425. Under the assumption that G3425 can form through the merger of the inner binary in the TCE process, we constrain the mass range as 2.5 M$_{\odot}$ $<$ $m_{\rm 1,ini} + m_{\rm 2,ini}$ $<$ 2.7 M$_{\odot}$, which roughly corresponds to the mass of the visible companions of G3425. Additionally, according to the discussion by \cite{2024NatAs.tmp..215W}, the BH mass that forms G3425 is estimated to require a helium core mass before the SN between 5 M$_{\odot}$ and 7 M$_{\odot}$. Therefore, we constrain the mass range as 21 M$_{\odot}$ $<$ $m_{\rm 3,ini}$ $<$ 22 M$_{\odot}$. On the other hand, the results of the 3D simulations of TCE by \cite{2021MNRAS.500.1921G} show that, during the TCE process, the inner binary can only inspiral faster than the outer binary if the orbital separation is relatively small (about 3$\sim$26 R$_{\odot}$ in their simulations). Therefore, we constrain the range as 5 R$_{\odot}$ $<$ $a_{\rm in,ini}$ $<$ 20 R$_{\odot}$. For the outer orbit, considering the possibility that the tertiary can fill its Roche lobe (with the maximum radius of about 1000$\sim$4000 R$_{\odot}$), we constrain $a_{\rm out,ini}$ $<$ 15 au. Combining this with the stability criterion for the initial triple, we perform Monte Carlo simulations within these ranges to search for orbits that resemble the formation of G3425. We selecte an initially dynamically stable triple, which undergoes TCE to evolve into a PTB, and eventually forms G3425 after a SN event. The initial masses of the three stars in the triple system are 1.49 $M_\odot$, 1.05 $M_\odot$, and 21.81 $M_\odot$; the inner and outer orbital periods are 4.22 days and 1961.78 days, respectively; and the inner and outer eccentricities are 0.08 and 0.15, respectively. Additionally, the inner and outer inclinations (radians), which are 1.73 and 1.27, respectively; the inner and outer arguments of pericentre (radians), which are 4.68 and 3.61, respectively; and the inner and outer longitudes of ascending node (radians), which are 4.35 and 0.61, respectively.

\begin{figure*}[htb]
\centering
\includegraphics[width=0.75\textwidth]{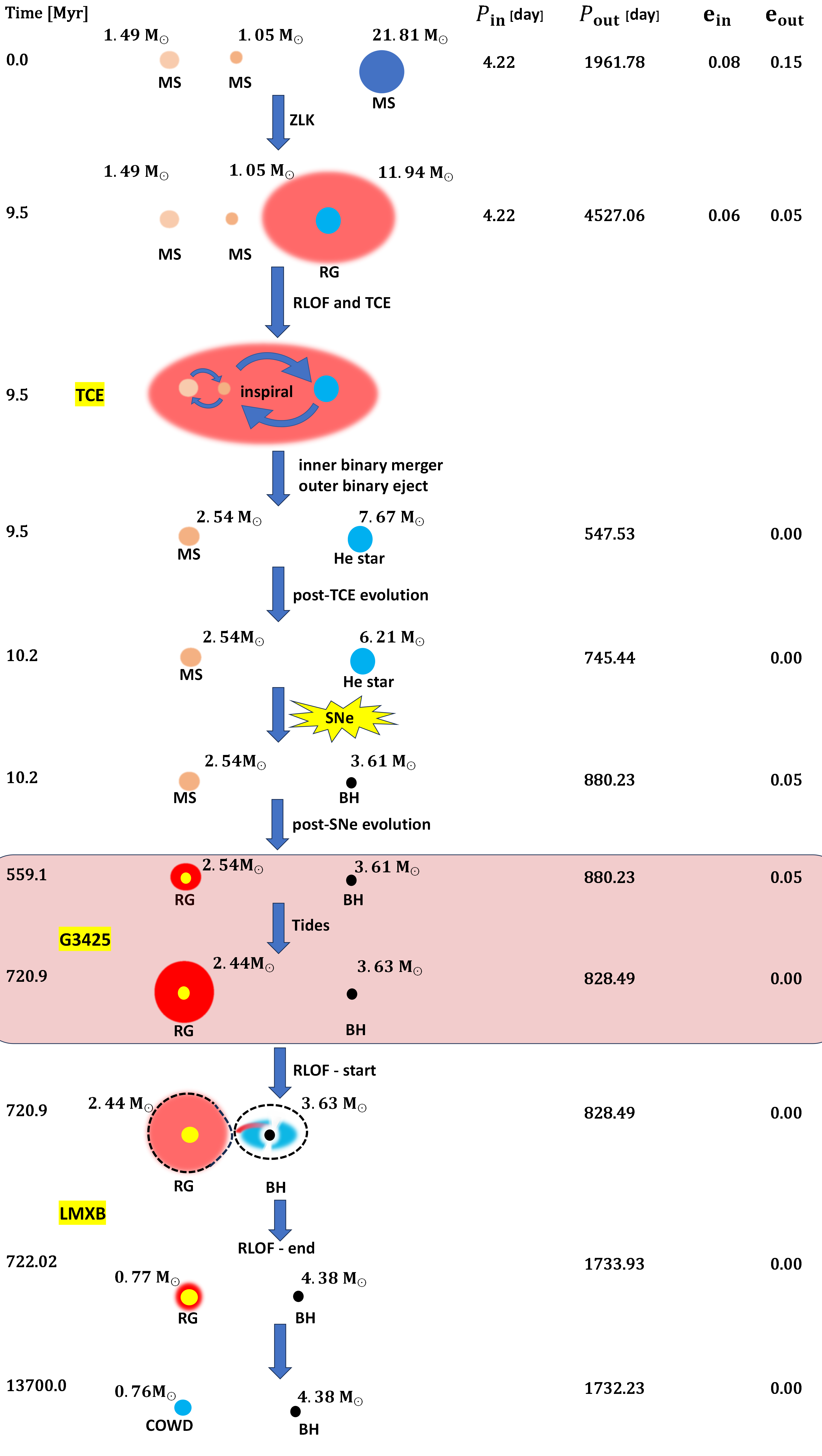}
\caption{The key evolutionary stages diagram of the initial triple leading to the formation of G3425, where the light red shaded area represents the duration of the observed G3425 phase.}
\label{fig:1}
\end{figure*}

\begin{figure*}[htb]
\centering
\includegraphics[width=\textwidth]{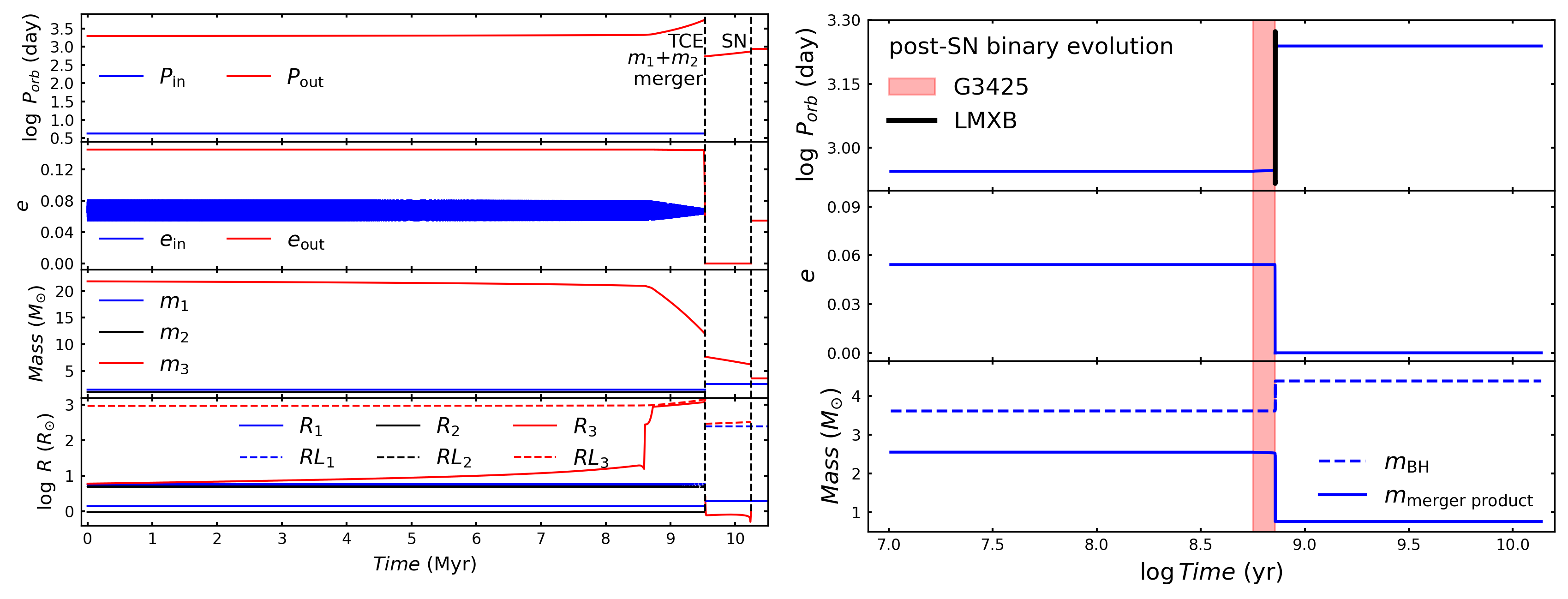}
\caption{The time functions of orbital period ($P_{\rm orb}$), eccentricity ($e$), mass, and radius for the selected triple.}
\label{fig:2}
\end{figure*}

In the Fig. \ref{fig:1} and Fig. \ref{fig:2}, we show the motion diagram at key points during the evolution of the selected triple, and the functions of orbital period, eccentricity, mass, and radius over time, respectively. In this triple, the tertiary has the greatest mass, evolves at the fastest rate, and its Roche lobe around the inner binary ($\sim$ 925 R$_{\odot}$). As the tertiary evolves, its radius begins to expand, and the outer orbital period evolves through adiabatic expansion due to mass loss from stellar winds, with $e_{\rm out}$, remaining nearly unchanged \citep{2002MNRAS.329..897H}. On the other hand, the ZLK effect, triggered by the exchange of angular momentum between the inner and outer orbits, causes $e_{\rm in}$ to undergo long-term oscillations \citep{1910AN....183..345V,1962AJ.....67..591K,1962P&SS....9..719L,2016ARA&A..54..441N} within the range of 0.05$\sim$0.08. At $\sim$8.6 Myr, the tertiary leaves MS stage, quickly passes through the Hertzsprung Gap (HG), and enters the RG phase. During this time, its radius rapidly expands, and a large amount of mass is lost through RG stellar winds. This results in a rapid increase in the $P_{\rm out}$ and strengthening of tidal \citep{2002MNRAS.329..897H}. The ZLK effect is suppressed by the short-range tidal forces, leading to a gradual reduction in the oscillation amplitude \citep{1997Natur.386..254H,2001ApJ...562.1012E,2002ApJ...578..775B,2017MNRAS.467.3066A,2024MNRAS.527.9782D}. At $\sim$9.5 Myr, the tertiary undergoes RLOF, with $P_{\rm in} = 4$ days ($a_{\rm in} = 15$ R$_{\odot}$), $P_{\rm out} = 4527$ days ($a_{\rm out} = 2807$ R$_{\odot}$), and $m_{3}=11.94$ M$_{\odot}$ ($m_{\rm 3,c}=7.67$ M$_{\odot}$ and $m_{\rm 3,env}=4.27$ M$_{\odot}$), and ($m_{1} + m_{2}$) = 2.54 M$_{\odot}$. Due to the very large mass ratio ($q = m_{3}/(m_{1}+m_{2})$ = 4.7), the MSE code determines that the MT is unstable, and the triple enters the TCE phase.

Based on the simulations by \cite{2021MNRAS.500.1921G} and the modeling in Section 2.2, we assume that the inner binary merges due to inspiral during the TCE phase. Combining Equations 2 and 3, we calculate the $E_{\rm bind}$ to be $3.75\times10^{48}$ ${\rm erg}$, and the separation of the inner binary from 15 R$_{\odot}$ of separation before inspiral to $(R_{1}+R_{2} )\cong$ 2.34 R$_{\odot}$ when merger occurs after inspiral produces $\sim$$3.68\times10^{48}$ ${\rm erg}$ of $E_{\rm orb,in}$, which is about 97\% of the $E_{\rm bind}$. This means that, in order to successfully eject CE, the remaining 3\% of the $E_{\rm bind}$ needs to be provided by the $E_{\rm orb,out}$. Under this assumption of $\alpha_{\rm CE,in}=\alpha_{\rm CE,out}=1$, we calculate that the $a_{\rm out}$ only needs to inspiral from its initial value of 2807 R$_{\odot}$ to 687 R$_{\odot}$ to successfully eject CE. It is worth noting that if a larger $\alpha_{\rm CE,in}$ or/ and $\alpha_{\rm CE,out}$ is assumed when the inner binary merges, we expect it to result in a smaller required $\Delta E_{\rm orb,out}$ for ejecting the CE, and the $a_{\rm out}$ after ejecting CE will be larger. In Fig. \ref{fig:1}, the merger product of the inner binary, along with the successfully ejected core of the donor, forms a PTB with an orbital separation (orbital period) of 687 R$_{\odot}$ (547 days) and an eccentricity of 0. We find that this result is very similar to the simulations by \cite{2021MNRAS.500.1921G}. Compared to traditional BCE, in TCE, the $E_{\rm orb}$ generated by the inspiral of the inner binary due to friction as an additional energy source transferred to CE. This accelerates the expansion and ejection of CE, thereby reducing the dissipation of $E_{\rm orb,out}$ and the inspiral of the outer orbit. This means that, without increasing $\alpha_{\rm ce}$, TCE is more likely than BCE to produce binaries with long $P_{\rm orb}$.

\begin{figure*}[htb]
\centering
\includegraphics[width=1\textwidth]{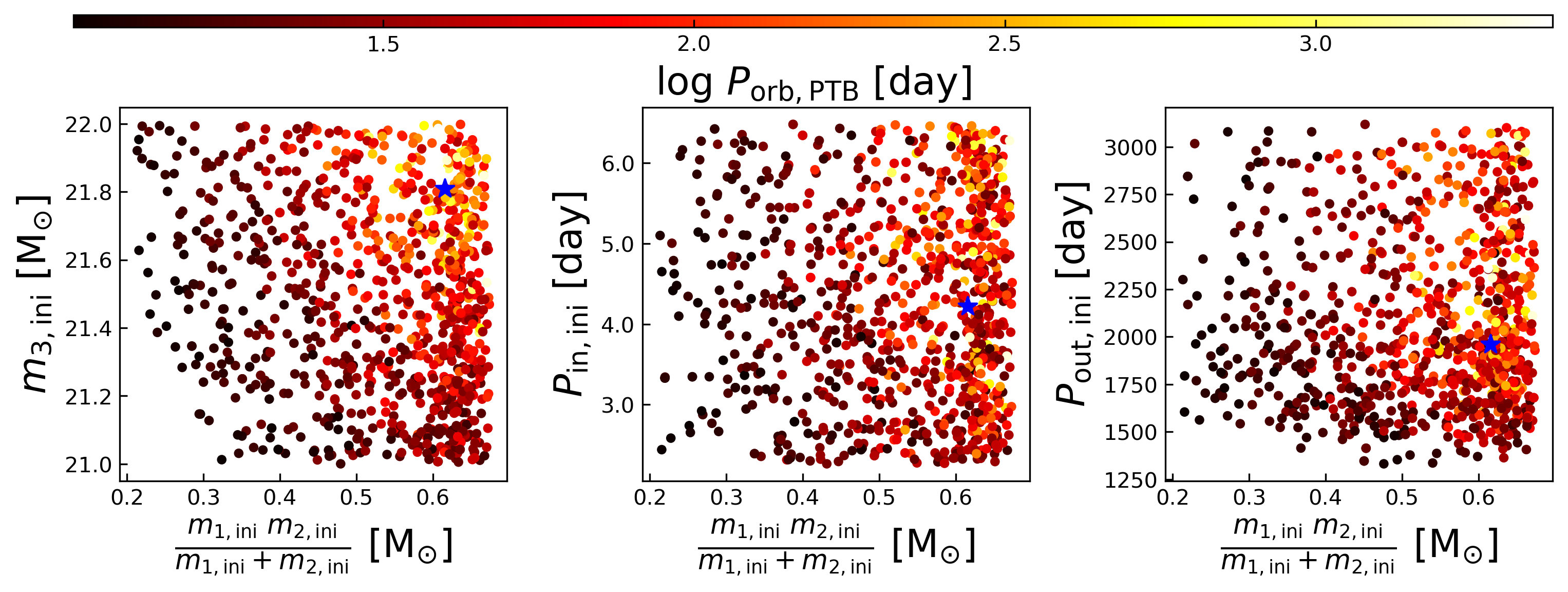}
\caption{The distribution of the initial reduced mass of the inner binary ($\frac{m_{\rm 1,ini} m_{\rm 2,ini}}{m_{\rm 1,ini} + m_{\rm 2,ini}}$), the initial mass of the tertiary, the initial inner and outer orbital periods, and the orbital period of the post-TCE binaries ($P_{\rm PTB,orb}$) under the assumption that $\alpha_{\rm CE,in}=\alpha_{\rm CE,out}=1$. Different colors represent the orbital periods of the PTBs, with the blue stars indicating the initial parameters selected in Fig. \ref{fig:1} and Fig. \ref{fig:2}.}
\label{fig:3}
\end{figure*}

Additionally, Fig. \ref{fig:3} shows the distribution of the initial masses of the three stars, the initial inner and outer orbital periods, and the orbital period of the post-TCE binaries ($P_{\rm orb,PTB}$) under the assumption that $\alpha_{\rm CE,in}=\alpha_{\rm CE,out}=1$. In our TCE model, the $E_{\rm bind}$ of ejecting the tertiary is mainly provided by $\Delta E_{\rm orb,in}$, which is primarily determined by $\frac{m_{\rm 1,ini} m_{\rm 2,ini}}{m_{\rm 1,ini} + m_{\rm 2,ini}}$ (see Equation 3). The larger $\frac{m_{\rm 1,ini} m_{\rm 2,ini}}{m_{\rm 1,ini} + m_{\rm 2,ini}}$ is, the greater the $\Delta E_{\rm orb,in}$, which ultimately increases the likelihood of forming PTBs with longer orbital periods (see Fig. \ref{fig:3}). In addition, the larger $m_{\rm 3,ini}$, the greater its mass-loss rate \citep{2001A&A...369..574V}. This results in a smaller envelope mass and a larger $a_{\rm out}$ at the time of TCE, and ultimately a smaller $E_{\rm bind}$ during the TCE process. This also increases the likelihood of forming PTBs with longer orbital periods (see the left panel of Fig. \ref{fig:3}).  On the other hand, $\Delta E_{\rm orb,in}$ and $\Delta E_{\rm orb,out}$ are mainly determined by the final orbital energy ($E_{\rm orb,in,f}$ and $E_{\rm orb,out,f}$), because the $a_{\rm in,i}$ and $a_{\rm out,i}$ are very large, and its $E_{\rm orb,in,i}$ and $E_{\rm orb,out,i}$ can be almost neglected (see Equation 3). This ultimately leads to a weak correlation between the size of the $P_{\rm orb,PTB}$ and $P_{\rm in,ini}$, $P_{\rm out,ini}$ (see the middle and right panels of Fig. \ref{fig:3}). Most of $P_{\rm orb,PTB}$ range from 10 to 100 days, followed by 100 to 1000 days, with only a very few $P_{\rm orb,PTB}$ greater than 1000 days (in our calculation, the maximum $P_{\rm orb,PTB}$ is $\sim$1850 days). We estimate that the $P_{\rm orb,PTB}$ is 1 to 2 orders of magnitude larger than those of the binaries formed through BCE evolution (under the same $\alpha_{\rm CE}$ assumption).

In the Fig. \ref{fig:1}, we use the binary module with a metallicity of Z = 0.014 of MESA to further track the evolution of PTB. Due to the strong Wolf-Rayet winds (at a mass loss rate of $\sim10^{-5.5} {\rm M_{\odot}/yr}$), $P_{\rm orb}$ widens to 745 days. Around 10.2 Myr, the helium star's core depletes carbon, at which point it is assumed to undergo a SN. Additionally, at the time of SN, the helium star has a mass of 6.21 M$_{\odot}$, with a $M_{\rm CO}$ of 4.21 M$_{\odot}$.

\begin{figure*}[htb]
\centering
\includegraphics[width=\textwidth]{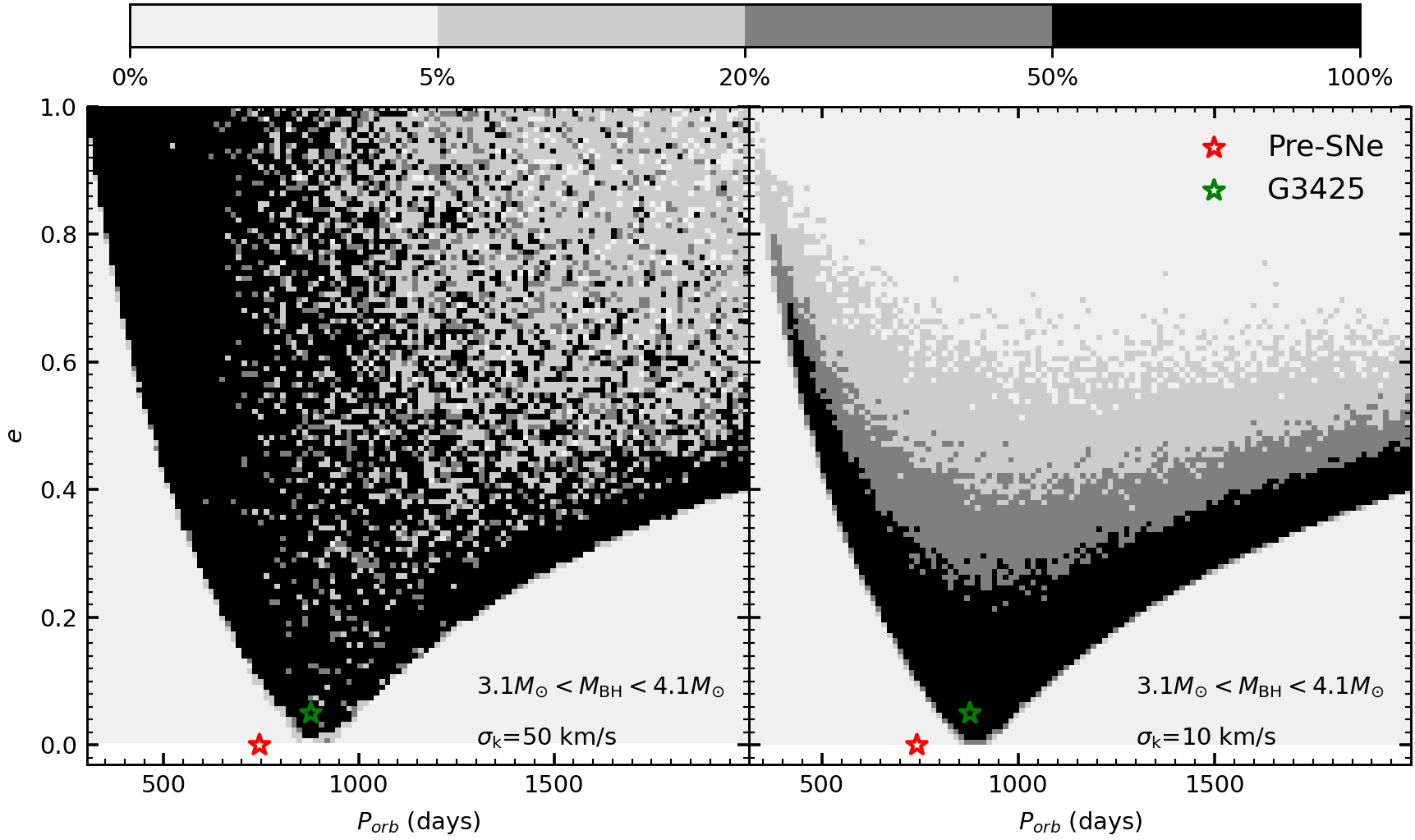}
\caption{The probability distribution of orbital period and eccentricity for the surviving BH binary of the PTB after undergoing SNe event. The darker the color, the higher the probability. The red and green markers represent the locations of the PTB before the SNe and the observed location of G3425, respectively. The left and right panels show the Maxwellian distributions for $\sigma_{\rm k}$ values of 50 ${\rm km/s}$ and 10 ${\rm km/s}$, respectively.}
\label{fig:4}
\end{figure*}

Using Equation 4 and Equation 5, we calculate the $P_{\rm BH}=\frac{4.21-2}{5}\cong$ 44.2\%, and the $P_{\rm CF}=\frac{4.21-2}{6}\cong$ 36.8\% in the SNe event. It is worth noting that, as described in Section 2.2, if the BH is formed through a complete fallback process, in our model, the resulting BH mass would be equal to the helium star mass ($\sim$ 6.21 M$_{\odot}$), which does not match the BH mass of G3425. In our calculation, the probability of forming a BH through this non-complete fallback process is $P=P_{\rm BH}(1-P_{\rm CF}) \cong 27.9\%$. We also calculate that in the case of non-CF, the remnant mass distribution follows a normal distribution with a mean of 3.37 M$_{\odot}$ and a standard deviation of 0.5 M$_{\odot}$. The BH mass in G3425 ($\sim$3.6 M$_{\odot}$), falls within one standard deviation from the mean of this normal distribution. Fig. \ref{fig:4} shows the probability distribution of the post-SN surviving BH binary orbits, for the systems selected in Fig. \ref{fig:1}, under different $\sigma_{\rm k}$ values for the Maxwellian distribution. Since $\sigma_{\rm k}$ = 50 ${\rm km/s}$ is larger than $\sigma_{\rm k}$ = 10 ${\rm km/s}$, the PTB typically receives a greater natal kick, resulting in a more dispersed orbital probability distribution and a lower survival rate for the PTB (survival rates of 8.5\% and 36.9\% for the left and right panels, respectively). In the Fig. \ref{fig:4}, by comparing with the observed orbital properties of G3425, we find that G3425 lies in the highest probability region (black area) of the surviving post-SNe BH binary orbit distribution, and the natal kick required to form G3425 are approximately $49^{+39}_{-39}$ ${\rm km/s}$ and $11^{+16}_{-5}$ ${\rm km/s}$, respectively. However, we find that forming G3425 from this SNe event remains quite challenging, as there is still a significant probability ($\sim$55.8\%) that the helium star would form a NS during the SNe event. Additionally, it is necessary to avoid forming a BH through CF, and finally, the PTB needs to avoid being disrupted during the SNe event. This conclusion is very similar to the simulation results by \cite{2024NatAs.tmp..215W}.

In the Fig. \ref{fig:1}, we select a post-SN BH binary orbit that is closest to G3425 and continue to track its evolution. At $\sim$ 559 Myr, the merger product of the inner binary leaves MS and enters the RG phase, indicating that the binary reaches the G3425 stage. The G3425 phase lasts for about 161.8 Myr. At around 720.9 Myr, the RG star fills its Roche lobe, marking the end of the G3425 phase and the beginning of the low mass X-ray binary (LMXB) phase. During MT process, the mass of donor is less than that of the accretor, causing the separation to gradually widen, and the Roche lobe radius to increase. Additionally, due to mass loss from the RG and the BH being constrained by the Eddington accretion rate, approximately 0.92 M$_{\odot}$ of mass is lost from the system during the MT process. When the Roche lobe radius becomes greater than the radius of the donor, MT stops. At this point, the $P_{\rm orb}$ is 1733 days, with the mass of $M_{\rm BH}$ being 4.38 M$_{\odot}$ and the RG having a mass of 0.77 M$_{\odot}$. The LMXB phase lasts approximately 1.12 Myr, during which the mass accretion rate of the BH ($\sim$ $6.69\times10^{-7}$ M$_{\odot}$/${\rm yr}$) is about 20 times higher than the Eddington accretion rate. Therefore, this LMXB phase is likely to be an ultraluminous X-ray source \citep{2020ApJ...902..125A}. In the subsequent evolution, the RG cools and forms a 0.76 M$_{\odot}$ carbon-oxygen white dwarf, resulting in a final $P_{\rm orb}$ of 1732.23 days, making it impossible for the system to merge within a Hubble time.

\section{Conclusion}
Using the MSE and MESA codes, we discuss that G3425 originated from a triple and evolved through a TCE. Based on the results of the 3D simulations of TCE by \cite{2021MNRAS.500.1921G}, we estimate the possible outcomes of TCE using the standard energy formalism on a 1D scale. We find that when the inner binary merges during the TCE process due to inspiral, the resulting $\Delta E_{\rm orb,in}$ contributes a significant proportion of $E_{\rm bind}$, approximately 97\% in our simulations. This means that, during TCE, the outer orbit does not need to supply much $\Delta E_{\rm orb,out}$ to successfully eject CE. Therefore, in our simulations, the final outcome of TCE is the merger of the inner binary, the successful ejection of the donor's core, and an ejected orbital separation that is 1-2 orders of magnitude larger than that of classical BCE with successful ejection. The result of this simulation is very similar to that of \cite{2021MNRAS.500.1921G}, despite our modeling being conducted on a 1D scale. It is worth noting that in our TCE simulation, we did not increase the $\alpha_{\rm CE}$, instead using $\alpha_{\rm CE}$ = 1. Subsequently, in the SNe simulations, we find that it is still challenging for the ejected helium star to form G3425, even though the observed G3425 falls within the highest probability region of the surviving post-SNe BH binary orbital distribution in our simulations. The reason is that this helium star not only has a significant probability ($\sim$55.8\%) of forming a NS after the SNe but also needs to avoid forming a BH through CF. Finally, the PTB must survive duting the SNe event, with survival rates of 8.5\% and 36.9\% for $\sigma_{\rm k}$ of 50 ${\rm km/s}$ and 10 ${\rm km/s}$, respectively. This could potentially explain why BHs in the mass gap are so rare. Additionally, in cases where $\sigma_{\rm k}$ is 50 ${\rm km/s}$ and 10 ${\rm km/s}$, the natal kicks required to form G3425 are $49^{+39}_{-39}$ ${\rm km/s}$ and $11^{+16}_{-5}$ ${\rm km/s}$, respectively. Combined with the calculated survival rates, these results suggest that the formation of G3425 is more likely to occur with a low natal kick.


\section*{Acknowledgements}
This work received the support of the National Natural Science Foundation of China under grants U2031204, 12163005, 12373038, and 12288102; the Natural Science Foundation of Xinjiang No.2022TSYCLJ0006 and 2022D01D85; the science research grants from the China Manned Space Project with No.CMS-CSST-2021-A10.

\bibliography{./sample631}

\begin{thebibliography}{}
\expandafter\ifx\csname natexlab\endcsname\relax\def\natexlab#1{#1}\fi
\providecommand{\url}[1]{\href{#1}{#1}}
\providecommand{\dodoi}[1]{doi:~\href{http://doi.org/#1}{\nolinkurl{#1}}}
\providecommand{\doeprint}[1]{\href{http://ascl.net/#1}{\nolinkurl{http://ascl.net/#1}}}
\providecommand{\doarXiv}[1]{\href{https://arxiv.org/abs/#1}{\nolinkurl{https://arxiv.org/abs/#1}}}

\bibitem[{{Abdusalam} {et~al.}(2020){Abdusalam}, {Ablimit}, {Hashim}, {L{\"u}},
  {Mardini}, \& {Wang}}]{2020ApJ...902..125A}
{Abdusalam}, K., {Ablimit}, I., {Hashim}, P., {et~al.} 2020, \apj, 902, 125,
  \dodoi{10.3847/1538-4357/abb5a8}

\bibitem[{{Anderson} {et~al.}(2017){Anderson}, {Lai}, \&
  {Storch}}]{2017MNRAS.467.3066A}
{Anderson}, K.~R., {Lai}, D., \& {Storch}, N.~I. 2017, \mnras, 467, 3066,
  \dodoi{10.1093/mnras/stx293}

\bibitem[{{Andrews} \& {Kalogera}(2022)}]{2022ApJ...930..159A}
{Andrews}, J.~J., \& {Kalogera}, V. 2022, \apj, 930, 159,
  \dodoi{10.3847/1538-4357/ac66d6}

\bibitem[{{Antonini} {et~al.}(2017){Antonini}, {Toonen}, \&
  {Hamers}}]{2017ApJ...841...77A}
{Antonini}, F., {Toonen}, S., \& {Hamers}, A.~S. 2017, \apj, 841, 77,
  \dodoi{10.3847/1538-4357/aa6f5e}

\bibitem[{{Bailyn} {et~al.}(1997){Bailyn}, {Jain}, {Coppi}, \&
  {Orosz}}]{1997AAS...190.1001B}
{Bailyn}, C.~D., {Jain}, R.~K., {Coppi}, P., \& {Orosz}, J.~A. 1997, in
  American Astronomical Society Meeting Abstracts, Vol. 190, American
  Astronomical Society Meeting Abstracts \#190, 10.01

\bibitem[{{Belczynski} \& {Ziolkowski}(2009)}]{2009ApJ...707..870B}
{Belczynski}, K., \& {Ziolkowski}, J. 2009, \apj, 707, 870,
  \dodoi{10.1088/0004-637X/707/2/870}

\bibitem[{{Blaes} {et~al.}(2002){Blaes}, {Lee}, \&
  {Socrates}}]{2002ApJ...578..775B}
{Blaes}, O., {Lee}, M.~H., \& {Socrates}, A. 2002, \apj, 578, 775,
  \dodoi{10.1086/342655}

\bibitem[{{Brott} {et~al.}(2011){Brott}, {de Mink}, {Cantiello}, {Langer}, {de
  Koter}, {Evans}, {Hunter}, {Trundle}, \& {Vink}}]{2011A&A...530A.115B}
{Brott}, I., {de Mink}, S.~E., {Cantiello}, M., {et~al.} 2011, \aap, 530, A115,
  \dodoi{10.1051/0004-6361/201016113}

\bibitem[{{Bruenech} {et~al.}(2024){Bruenech}, {Boekholt}, {Kummer}, \&
  {Toonen}}]{2024arXiv240811128B}
{Bruenech}, C.~W., {Boekholt}, T., {Kummer}, F., \& {Toonen}, S. 2024, arXiv
  e-prints, arXiv:2408.11128, \dodoi{10.48550/arXiv.2408.11128}

\bibitem[{{Burdge} {et~al.}(2024){Burdge}, {El-Badry}, {Kara}, {Canizares},
  {Chakrabarty}, {Frebel}, {Millholland}, {Rappaport}, {Simcoe}, \&
  {Vanderburg}}]{2024Natur.635..316B}
{Burdge}, K.~B., {El-Badry}, K., {Kara}, E., {et~al.} 2024, \nat, 635, 316,
  \dodoi{10.1038/s41586-024-08120-6}

\bibitem[{{Burrows} {et~al.}(2023){Burrows}, {Vartanyan}, \&
  {Wang}}]{2023ApJ...957...68B}
{Burrows}, A., {Vartanyan}, D., \& {Wang}, T. 2023, \apj, 957, 68,
  \dodoi{10.3847/1538-4357/acfc1c}

\bibitem[{{Casares} \& {Jonker}(2014)}]{2014SSRv..183..223C}
{Casares}, J., \& {Jonker}, P.~G. 2014, \ssr, 183, 223,
  \dodoi{10.1007/s11214-013-0030-6}

\bibitem[{{Chakrabarti} {et~al.}(2023){Chakrabarti}, {Simon}, {Craig},
  {Reggiani}, {Brandt}, {Guhathakurta}, {Dalba}, {Kirby}, {Chang}, {Hey},
  {Savino}, {Geha}, \& {Thompson}}]{2023AJ....166....6C}
{Chakrabarti}, S., {Simon}, J.~D., {Craig}, P.~A., {et~al.} 2023, \aj, 166, 6,
  \dodoi{10.3847/1538-3881/accf21}

\bibitem[{{Chawla} {et~al.}(2022){Chawla}, {Chatterjee}, {Breivik}, {Moorthy},
  {Andrews}, \& {Sanderson}}]{2022ApJ...931..107C}
{Chawla}, C., {Chatterjee}, S., {Breivik}, K., {et~al.} 2022, \apj, 931, 107,
  \dodoi{10.3847/1538-4357/ac60a5}

\bibitem[{{Chawla} {et~al.}(2024){Chawla}, {Chatterjee}, {Shah}, \&
  {Breivik}}]{2024ApJ...975..163C}
{Chawla}, C., {Chatterjee}, S., {Shah}, N., \& {Breivik}, K. 2024, \apj, 975,
  163, \dodoi{10.3847/1538-4357/ad7b0b}

\bibitem[{{Coleman} \& {Burrows}(2022)}]{2022MNRAS.517.3938C}
{Coleman}, M. S.~B., \& {Burrows}, A. 2022, \mnras, 517, 3938,
  \dodoi{10.1093/mnras/stac2573}

\bibitem[{{Comerford} \& {Izzard}(2020)}]{2020MNRAS.498.2957C}
{Comerford}, T.~A.~F., \& {Izzard}, R.~G. 2020, \mnras, 498, 2957,
  \dodoi{10.1093/mnras/staa2539}

\bibitem[{{Corral-Santana} {et~al.}(2016){Corral-Santana}, {Casares},
  {Mu{\~n}oz-Darias}, {Bauer}, {Mart{\'\i}nez-Pais}, \&
  {Russell}}]{2016A&A...587A..61C}
{Corral-Santana}, J.~M., {Casares}, J., {Mu{\~n}oz-Darias}, T., {et~al.} 2016,
  \aap, 587, A61, \dodoi{10.1051/0004-6361/201527130}

\bibitem[{{De} {et~al.}(2024){De}, {MacLeod}, {Jencson}, {Lovegrove}, {Antoni},
  {Kara}, {Kasliwal}, {Lau}, {Loeb}, {Masterson}, {Meisner}, {Panagiotou},
  {Quataert}, \& {Simcoe}}]{2024arXiv241014778D}
{De}, K., {MacLeod}, M., {Jencson}, J.~E., {et~al.} 2024, arXiv e-prints,
  arXiv:2410.14778, \dodoi{10.48550/arXiv.2410.14778}

\bibitem[{{de Vries} {et~al.}(2014){de Vries}, {Portegies Zwart}, \&
  {Figueira}}]{2014MNRAS.438.1909D}
{de Vries}, N., {Portegies Zwart}, S., \& {Figueira}, J. 2014, \mnras, 438,
  1909, \dodoi{10.1093/mnras/stt1688}

\bibitem[{{Dorozsmai} {et~al.}(2024){Dorozsmai}, {Toonen}, {Vigna-G{\'o}mez},
  {de Mink}, \& {Kummer}}]{2024MNRAS.527.9782D}
{Dorozsmai}, A., {Toonen}, S., {Vigna-G{\'o}mez}, A., {de Mink}, S.~E., \&
  {Kummer}, F. 2024, \mnras, 527, 9782, \dodoi{10.1093/mnras/stad3819}

\bibitem[{{Eggleton}(1983)}]{1983ApJ...268..368E}
{Eggleton}, P.~P. 1983, \apj, 268, 368, \dodoi{10.1086/160960}

\bibitem[{{Eggleton} \& {Kiseleva-Eggleton}(2001)}]{2001ApJ...562.1012E}
{Eggleton}, P.~P., \& {Kiseleva-Eggleton}, L. 2001, \apj, 562, 1012,
  \dodoi{10.1086/323843}

\bibitem[{{Ekstr{\"o}m} {et~al.}(2012){Ekstr{\"o}m}, {Georgy}, {Eggenberger},
  {Meynet}, {Mowlavi}, {Wyttenbach}, {Granada}, {Decressin}, {Hirschi},
  {Frischknecht}, {Charbonnel}, \& {Maeder}}]{2012A&A...537A.146E}
{Ekstr{\"o}m}, S., {Georgy}, C., {Eggenberger}, P., {et~al.} 2012, \aap, 537,
  A146, \dodoi{10.1051/0004-6361/201117751}

\bibitem[{{El-Badry} {et~al.}(2023{\natexlab{a}}){El-Badry}, {Rix}, {Quataert},
  {Howard}, {Isaacson}, {Fuller}, {Hawkins}, {Breivik}, {Wong}, {Rodriguez},
  {Conroy}, {Shahaf}, {Mazeh}, {Arenou}, {Burdge}, {Bashi}, {Faigler}, {Weisz},
  {Seeburger}, {Almada Monter}, \& {Wojno}}]{2023MNRAS.518.1057E}
{El-Badry}, K., {Rix}, H.-W., {Quataert}, E., {et~al.} 2023{\natexlab{a}},
  \mnras, 518, 1057, \dodoi{10.1093/mnras/stac3140}

\bibitem[{{El-Badry} {et~al.}(2023{\natexlab{b}}){El-Badry}, {Rix}, {Cendes},
  {Rodriguez}, {Conroy}, {Quataert}, {Hawkins}, {Zari}, {Hobson}, {Breivik},
  {Rau}, {Berger}, {Shahaf}, {Seeburger}, {Burdge}, {Latham}, {Buchhave},
  {Bieryla}, {Bashi}, {Mazeh}, \& {Faigler}}]{2023MNRAS.521.4323E}
{El-Badry}, K., {Rix}, H.-W., {Cendes}, Y., {et~al.} 2023{\natexlab{b}},
  \mnras, 521, 4323, \dodoi{10.1093/mnras/stad799}

\bibitem[{{Eldridge} {et~al.}(2017){Eldridge}, {Stanway}, {Xiao}, {McClelland},
  {Taylor}, {Ng}, {Greis}, \& {Bray}}]{2017PASA...34...58E}
{Eldridge}, J.~J., {Stanway}, E.~R., {Xiao}, L., {et~al.} 2017, \pasa, 34,
  e058, \dodoi{10.1017/pasa.2017.51}

\bibitem[{{Farr} {et~al.}(2011){Farr}, {Sravan}, {Cantrell}, {Kreidberg},
  {Bailyn}, {Mandel}, \& {Kalogera}}]{2011ApJ...741..103F}
{Farr}, W.~M., {Sravan}, N., {Cantrell}, A., {et~al.} 2011, \apj, 741, 103,
  \dodoi{10.1088/0004-637X/741/2/103}

\bibitem[{{Fragos} {et~al.}(2023){Fragos}, {Andrews}, {Bavera}, {Berry},
  {Coughlin}, {Dotter}, {Giri}, {Kalogera}, {Katsaggelos}, {Kovlakas},
  {Lalvani}, {Misra}, {Srivastava}, {Qin}, {Rocha}, {Rom{\'a}n-Garza}, {Serra},
  {Stahle}, {Sun}, {Teng}, {Trajcevski}, {Tran}, {Xing}, {Zapartas}, \&
  {Zevin}}]{2023ApJS..264...45F}
{Fragos}, T., {Andrews}, J.~J., {Bavera}, S.~S., {et~al.} 2023, \apjs, 264, 45,
  \dodoi{10.3847/1538-4365/ac90c1}

\bibitem[{{Fryer} {et~al.}(2012){Fryer}, {Belczynski}, {Wiktorowicz},
  {Dominik}, {Kalogera}, \& {Holz}}]{2012ApJ...749...91F}
{Fryer}, C.~L., {Belczynski}, K., {Wiktorowicz}, G., {et~al.} 2012, \apj, 749,
  91, \dodoi{10.1088/0004-637X/749/1/91}

\bibitem[{{Fryer} \& {Kalogera}(2001)}]{2001ApJ...554..548F}
{Fryer}, C.~L., \& {Kalogera}, V. 2001, \apj, 554, 548, \dodoi{10.1086/321359}

\bibitem[{{Gaia Collaboration} {et~al.}(2024){Gaia Collaboration}, {Panuzzo},
  {Mazeh}, {Arenou}, {Holl}, {Caffau}, {Jorissen}, {Babusiaux}, {Gavras},
  {Sahlmann}, {Bastian}, {Wyrzykowski}, {Eyer}, {Leclerc}, {Bauchet},
  {Bombrun}, {Mowlavi}, {Seabroke}, {Teyssier}, {Balbinot}, {Helmi}, {Brown},
  {Vallenari}, {Prusti}, {de Bruijne}, {Barbier}, {Biermann}, {Creevey},
  {Ducourant}, {Evans}, {Guerra}, {Hutton}, {Jordi}, {Klioner}, {Lammers},
  {Lindegren}, {Luri}, {Mignard}, {Nicolas}, {Randich}, {Sartoretti},
  {Smiljanic}, {Tanga}, {Walton}, {Aerts}, {Bailer-Jones}, {Cropper},
  {Drimmel}, {Jansen}, {Katz}, {Lattanzi}, {Soubiran}, {Th{\'e}venin}, {van
  Leeuwen}, {Andrae}, {Audard}, {Bakker}, {Blomme}, {Casta{\~n}eda}, {De
  Angeli}, {Fabricius}, {Fouesneau}, {Fr{\'e}mat}, {Galluccio}, {Guerrier},
  {Heiter}, {Masana}, {Messineo}, {Nienartowicz}, {Pailler}, {Riclet}, {Roux},
  {Sordo}, {Gracia-Abril}, {Portell}, {Altmann}, {Benson}, {Berthier},
  {Burgess}, {Busonero}, {Busso}, {Cacciari}, {C{\'a}novas}, {Carrasco},
  {Carry}, {Cellino}, {Cheek}, {Clementini}, {Damerdji}, {Davidson}, {de
  Teodoro}, {Delchambre}, {Dell'Oro}, {Fraile Garcia}, {Garabato},
  {Garc{\'\i}a-Lario}, {Haigron}, {Hambly}, {Harrison}, {Hatzidimitriou},
  {Hern{\'a}ndez}, {Hestroffer}, {Hodgkin}, {Jamal}, {Jevardat de Fombelle},
  {Jordan}, {Krone-Martins}, {Lanzafame}, {L{\"o}ffler}, {Lorca}, {Marchal},
  {Marrese}, {Moitinho}, {Muinonen}, {Nu{\~n}ez Campos}, {Oreshina-Slezak},
  {Osborne}, {Pancino}, {Pauwels}, {Recio-Blanco}, {Riello}, {Rimoldini},
  {Robin}, {Roegiers}, {Sarro}, {Schultheis}, {Smith}, {Sozzetti}, {Utrilla},
  {van Leeuwen}, {Weingrill}, {Abbas}, {{\'A}brah{\'a}m}, {Abreu Aramburu},
  {Ahmed}, {Altavilla}, {{\'A}lvarez}, {Anders}, {Anderson}, {Anglada Varela},
  {Antoja}, {Baig}, {Baines}, {Baker}, {Balaguer-N{\'u}{\~n}ez}, {Balog},
  {Barache}, {Barros}, {Barstow}, {Bartolom{\'e}}, {Bashi}, {Bassilana},
  {Baudeau}, {Becciani}, {Bedin}, {Bellas-Velidis}, {Bellazzini}, {Beordo},
  {Bernet}, {Bertolotto}, {Bertone}, {Bianchi}, {Binnenfeld},
  {Blanco-Cuaresma}, {Bland-Hawthorn}, {Blazere}, {Boch}, {Bossini},
  {Bouquillon}, {Bragaglia}, {Braine}, {Bratsolis}, {Breedt}, {Bressan},
  {Brouillet}, {Brugaletta}, {Bucciarelli}, {Butkevich}, {Buzzi}, {Camut},
  {Cancelliere}, {Cantat-Gaudin}, {Capilla Guilarte}, {Carballo}, {Carlucci},
  {Carnerero}, {Carretero}, {Carton}, {Casamiquela}, {Casey}, {Castellani},
  {Castro-Ginard}, {Ceraj}, {Cesare}, {Charlot}, {Chaudet}, {Chemin},
  {Chiavassa}, {Chornay}, {Chosson}, {Cooper}, {Cornez}, {Cowell}, {Crosta},
  {Crowley}, {Cruz Reyes}, {Dafonte}, {Dal Ponte}, {David}, {de Laverny}, {De
  Luise}, {De March}, {de Torres}, {del Peloso}, {Delbo}, {Delgado}, {Delisle},
  {Demouchy}, {Denis}, {Dharmawardena}, {Di Giacomo}, {Diener}, {Distefano},
  {Dolding}, {Dsilva}, {Enke}, {Fabre}, {Fabrizio}, {Faigler}, {Fatovi{\'c}},
  {Fedorets}, {Fern{\'a}ndez-Hern{\'a}ndez}, {Fernique}, {Figueras}, {Fouron},
  {Fragkoudi}, {Gai}, {Galinier}, {Garcia-Serrano}, {Garc{\'\i}a-Torres},
  {Garofalo}, {Gerlach}, {Geyer}, {Giacobbe}, {Gilmore}, {Girona}, {Giuffrida},
  {Gomboc}, {Gomez}, {Gonz{\'a}lez-Santamar{\'\i}a}, {Gosset}, {Granvik},
  {Gregori Barrera}, {Guti{\'e}rrez-S{\'a}nchez}, {Haywood}, {Helmer},
  {Hidalgo}, {Hilger}, {Hobbs}, {Hottier}, {Huckle}, {Jim{\'e}nez-Arranz},
  {Juaristi Campillo}, {Kaczmarek}, {Kervella}, {Khanna}, {Kontizas},
  {Kordopatis}, {Korn}, {K{\'o}sp{\'a}l}, {Kostrzewa-Rutkowska},
  {Kruszy{\'n}ska}, {Kun}, {Lambert}, {Lanza}, {Lebreton}, {Lebzelter},
  {Leccia}, {Lecoutre}, {Liao}, {Liberato}, {Licata}, {Livanou}, {Lobel},
  {L{\'o}pez-Miralles}, {Loup}, {Madar{\'a}sz}, {Mahy}, {Mann}, {Manteiga},
  {Marcellino}, {Marchant}, {Marconi}, {Mar{\'\i}n Pina}, {Marinoni},
  {Marshall}, {Mart{\'\i}n Lozano}, {Martin Polo}, {Mart{\'\i}n-Fleitas},
  {Marton}, {Mascarenhas}, {Masip}, {Mastrobuono-Battisti}, {McMillan},
  {Meichsner}, {Merc}, {Messina}, {Millar}, {Mints}, {Mohamed}, {Molina},
  {Molinaro}, {Moln{\'a}r}, {Mongui{\'o}}, {Montegriffo}, {Monti}, {Mora},
  {Morbidelli}, {Morris}, {Mudimadugula}, {Muraveva}, {Musella}, {Nagy},
  {Nardetto}, {Navarrete}, {Oh}, {Ordenovic}, {Orenstein}, {Pagani}, {Pagano},
  {Palaversa}, {Palicio}, {Pallas-Quintela}, {Pawlak}, {Penttil{\"a}},
  {Pesciullesi}, {Pinamonti}, {Plachy}, {Planquart}, {Plum}, {Poggio},
  {Pourbaix}, {Price-Whelan}, {Pulone}, {Rabin}, {Rainer}, {Raiteri}, {Ramos},
  {Ramos-Lerate}, {Ratajczak}, {Re Fiorentin}, {Regibo}, {Reyl{\'e}}, {Ripepi},
  {Riva}, {Rix}, {Rixon}, {Robert}, {Robichon}, {Robin}, {Romero-G{\'o}mez},
  {Rowell}, {Ruz Mieres}, {Rybicki}, {Sadowski}, {Sagrist{\`a} Sell{\'e}s},
  {Sanna}, {Santove{\~n}a}, {Sarasso}, {Sarmiento}, {Sarrate Riera}, {Sciacca},
  {S{\'e}gransan}, {Semczuk}, {Shahaf}, {Siebert}, {Slezak}, {Smart}, {Snaith},
  {Solano}, {Solitro}, {Souami}, {Souchay}, {Spitoni}, {Spoto}, {Squillante},
  {Steele}, {Steidelm{\"u}ller}, {Surdej}, {Szabados}, {Taris}, {Taylor},
  {Teixeira}, {Tepper-Garcia}, {Thuillot}, {Tolomei}, {Tonello}, {Torra},
  {Torralba Elipe}, {Trabucchi}, {Trentin}, {Tsantaki}, {Turon}, {Ulla},
  {Unger}, {Valtchanov}, {Vanel}, {Vecchiato}, {Vicente}, {Villar}, {Weiler},
  {Zhao}, {Zorec}, {Zucker}, {{\v{Z}}upi{\'c}}, \&
  {Zwitter}}]{2024A&A...686L...2G}
{Gaia Collaboration}, {Panuzzo}, P., {Mazeh}, T., {et~al.} 2024, \aap, 686, L2,
  \dodoi{10.1051/0004-6361/202449763}

\bibitem[{{Giacobbo} {et~al.}(2018){Giacobbo}, {Mapelli}, \&
  {Spera}}]{2018MNRAS.474.2959G}
{Giacobbo}, N., {Mapelli}, M., \& {Spera}, M. 2018, \mnras, 474, 2959,
  \dodoi{10.1093/mnras/stx2933}

\bibitem[{{Gilkis} \& {Mazeh}(2024)}]{2024MNRAS.535L..44G}
{Gilkis}, A., \& {Mazeh}, T. 2024, \mnras, 535, L44,
  \dodoi{10.1093/mnrasl/slae091}

\bibitem[{{Glanz} \& {Perets}(2021)}]{2021MNRAS.500.1921G}
{Glanz}, H., \& {Perets}, H.~B. 2021, \mnras, 500, 1921,
  \dodoi{10.1093/mnras/staa3242}

\bibitem[{{Hamers}(2018)}]{2018MNRAS.476.4139H}
{Hamers}, A.~S. 2018, \mnras, 476, 4139, \dodoi{10.1093/mnras/sty428}

\bibitem[{{Hamers}(2020)}]{2020MNRAS.494.5492H}
---. 2020, \mnras, 494, 5492, \dodoi{10.1093/mnras/staa1084}

\bibitem[{{Hamers} {et~al.}(2022{\natexlab{a}}){Hamers}, {Glanz}, \&
  {Neunteufel}}]{2022ApJS..259...25H}
{Hamers}, A.~S., {Glanz}, H., \& {Neunteufel}, P. 2022{\natexlab{a}}, \apjs,
  259, 25, \dodoi{10.3847/1538-4365/ac49e7}

\bibitem[{{Hamers} {et~al.}(2022{\natexlab{b}}){Hamers}, {Perets}, {Thompson},
  \& {Neunteufel}}]{2022ApJ...925..178H}
{Hamers}, A.~S., {Perets}, H.~B., {Thompson}, T.~A., \& {Neunteufel}, P.
  2022{\natexlab{b}}, \apj, 925, 178, \dodoi{10.3847/1538-4357/ac400b}

\bibitem[{{Hamers} \& {Portegies Zwart}(2016)}]{2016MNRAS.459.2827H}
{Hamers}, A.~S., \& {Portegies Zwart}, S.~F. 2016, \mnras, 459, 2827,
  \dodoi{10.1093/mnras/stw784}

\bibitem[{{Hamers} {et~al.}(2021){Hamers}, {Rantala}, {Neunteufel}, {Preece},
  \& {Vynatheya}}]{2021MNRAS.502.4479H}
{Hamers}, A.~S., {Rantala}, A., {Neunteufel}, P., {Preece}, H., \& {Vynatheya},
  P. 2021, \mnras, 502, 4479, \dodoi{10.1093/mnras/stab287}

\bibitem[{{Holman} {et~al.}(1997){Holman}, {Touma}, \&
  {Tremaine}}]{1997Natur.386..254H}
{Holman}, M., {Touma}, J., \& {Tremaine}, S. 1997, \nat, 386, 254,
  \dodoi{10.1038/386254a0}

\bibitem[{{Hurley} {et~al.}(2000){Hurley}, {Pols}, \&
  {Tout}}]{2000MNRAS.315..543H}
{Hurley}, J.~R., {Pols}, O.~R., \& {Tout}, C.~A. 2000, \mnras, 315, 543,
  \dodoi{10.1046/j.1365-8711.2000.03426.x}

\bibitem[{{Hurley} {et~al.}(2002){Hurley}, {Tout}, \&
  {Pols}}]{2002MNRAS.329..897H}
{Hurley}, J.~R., {Tout}, C.~A., \& {Pols}, O.~R. 2002, \mnras, 329, 897,
  \dodoi{10.1046/j.1365-8711.2002.05038.x}

\bibitem[{{Iben} \& {Livio}(1993)}]{1993PASP..105.1373I}
{Iben}, Icko, J., \& {Livio}, M. 1993, \pasp, 105, 1373, \dodoi{10.1086/133321}

\bibitem[{{Ivanova} {et~al.}(2020){Ivanova}, {Justham}, \&
  {Ricker}}]{2020cee..book.....I}
{Ivanova}, N., {Justham}, S., \& {Ricker}, P. 2020, {Common Envelope
  Evolution}, \dodoi{10.1088/2514-3433/abb6f0}

\bibitem[{{Ivanova} {et~al.}(2013){Ivanova}, {Justham}, {Chen}, {De Marco},
  {Fryer}, {Gaburov}, {Ge}, {Glebbeek}, {Han}, {Li}, {Lu}, {Marsh},
  {Podsiadlowski}, {Potter}, {Soker}, {Taam}, {Tauris}, {van den Heuvel}, \&
  {Webbink}}]{2013A&ARv..21...59I}
{Ivanova}, N., {Justham}, S., {Chen}, X., {et~al.} 2013, \aapr, 21, 59,
  \dodoi{10.1007/s00159-013-0059-2}

\bibitem[{{Janka} \& {Kresse}(2024)}]{2024Ap&SS.369...80J}
{Janka}, H.-T., \& {Kresse}, D. 2024, \apss, 369, 80,
  \dodoi{10.1007/s10509-024-04343-1}

\bibitem[{{Kimball} {et~al.}(2023){Kimball}, {Imperato}, {Kalogera}, {Rocha},
  {Doctor}, {Andrews}, {Dotter}, {Zapartas}, {Bavera}, {Kovlakas}, {Fragos},
  {Srivastava}, {Misra}, {Sun}, \& {Xing}}]{2023ApJ...952L..34K}
{Kimball}, C., {Imperato}, S., {Kalogera}, V., {et~al.} 2023, \apjl, 952, L34,
  \dodoi{10.3847/2041-8213/ace526}

\bibitem[{{Kippenhahn} \& {Weigert}(1994)}]{1994sse..book.....K}
{Kippenhahn}, R., \& {Weigert}, A. 1994, {Stellar Structure and Evolution}

\bibitem[{{Kotko} {et~al.}(2024){Kotko}, {Banerjee}, \&
  {Belczynski}}]{2024arXiv240313579K}
{Kotko}, I., {Banerjee}, S., \& {Belczynski}, K. 2024, arXiv e-prints,
  arXiv:2403.13579, \dodoi{10.48550/arXiv.2403.13579}

\bibitem[{{Kozai}(1962)}]{1962AJ.....67..591K}
{Kozai}, Y. 1962, \aj, 67, 591, \dodoi{10.1086/108790}

\bibitem[{{Kruckow} {et~al.}(2018){Kruckow}, {Tauris}, {Langer}, {Kramer}, \&
  {Izzard}}]{2018MNRAS.481.1908K}
{Kruckow}, M.~U., {Tauris}, T.~M., {Langer}, N., {Kramer}, M., \& {Izzard},
  R.~G. 2018, \mnras, 481, 1908, \dodoi{10.1093/mnras/sty2190}

\bibitem[{{Kruckow} {et~al.}(2024){Kruckow}, {Andrews}, {Fragos}, {Holl},
  {Bavera}, {Briel}, {Gossage}, {Kovlakas}, {Rocha}, {Sun}, {Srivastava},
  {Xing}, \& {Zapartas}}]{2024arXiv241018501K}
{Kruckow}, M.~U., {Andrews}, J.~J., {Fragos}, T., {et~al.} 2024, arXiv
  e-prints, arXiv:2410.18501, \dodoi{10.48550/arXiv.2410.18501}

\bibitem[{{Kummer} {et~al.}(2024){Kummer}, {Toonen}, {Dorozsmai}, {Grishin}, \&
  {de Koter}}]{2024arXiv240903826K}
{Kummer}, F., {Toonen}, S., {Dorozsmai}, A., {Grishin}, E., \& {de Koter}, A.
  2024, arXiv e-prints, arXiv:2409.03826, \dodoi{10.48550/arXiv.2409.03826}

\bibitem[{{Langer}(1991)}]{1991A&A...252..669L}
{Langer}, N. 1991, \aap, 252, 669

\bibitem[{{Li} {et~al.}(2024{\natexlab{a}}){Li}, {Zhu}, {L{\"u}}, {Li}, {Liu},
  {Guo}, {Yu}, \& {Lu}}]{2024ApJ...969..160L}
{Li}, Z., {Zhu}, C., {L{\"u}}, G., {et~al.} 2024{\natexlab{a}}, \apj, 969, 160,
  \dodoi{10.3847/1538-4357/ad4da8}

\bibitem[{{Li} {et~al.}(2024{\natexlab{b}}){Li}, {Zhu}, {Lu}, {L{\"u}}, {Li},
  {Liu}, {Guo}, \& {Yu}}]{2024ApJ...975L...8L}
{Li}, Z., {Zhu}, C., {Lu}, X., {et~al.} 2024{\natexlab{b}}, \apjl, 975, L8,
  \dodoi{10.3847/2041-8213/ad8653}

\bibitem[{{Lidov}(1962)}]{1962P&SS....9..719L}
{Lidov}, M.~L. 1962, \planss, 9, 719, \dodoi{10.1016/0032-0633(62)90129-0}

\bibitem[{{Livio} \& {Soker}(1988)}]{1988ApJ...329..764L}
{Livio}, M., \& {Soker}, N. 1988, \apj, 329, 764, \dodoi{10.1086/166419}

\bibitem[{{Loveridge} {et~al.}(2011){Loveridge}, {van der Sluys}, \&
  {Kalogera}}]{2011ApJ...743...49L}
{Loveridge}, A.~J., {van der Sluys}, M.~V., \& {Kalogera}, V. 2011, \apj, 743,
  49, \dodoi{10.1088/0004-637X/743/1/49}

\bibitem[{{Lu} {et~al.}(2023){Lu}, {Zhu}, {Liu}, {Guo}, {Yu}, \&
  {L{\"u}}}]{2023A&A...674A.216L}
{Lu}, X., {Zhu}, C., {Liu}, H., {et~al.} 2023, \aap, 674, A216,
  \dodoi{10.1051/0004-6361/202243188}

\bibitem[{{Mandel} \& {M{\"u}ller}(2020)}]{2020MNRAS.499.3214M}
{Mandel}, I., \& {M{\"u}ller}, B. 2020, \mnras, 499, 3214,
  \dodoi{10.1093/mnras/staa3043}

\bibitem[{{Mapelli} \& {Giacobbo}(2018)}]{2018MNRAS.479.4391M}
{Mapelli}, M., \& {Giacobbo}, N. 2018, \mnras, 479, 4391,
  \dodoi{10.1093/mnras/sty1613}

\bibitem[{{Mardling} \& {Aarseth}(2001)}]{2001MNRAS.321..398M}
{Mardling}, R.~A., \& {Aarseth}, S.~J. 2001, \mnras, 321, 398,
  \dodoi{10.1046/j.1365-8711.2001.03974.x}

\bibitem[{{Martinez} {et~al.}(2020){Martinez}, {Fragione}, {Kremer},
  {Chatterjee}, {Rodriguez}, {Samsing}, {Ye}, {Weatherford}, {Zevin}, {Naoz},
  \& {Rasio}}]{2020ApJ...903...67M}
{Martinez}, M. A.~S., {Fragione}, G., {Kremer}, K., {et~al.} 2020, \apj, 903,
  67, \dodoi{10.3847/1538-4357/abba25}

\bibitem[{{Mata Sanchez} {et~al.}(2024){Mata Sanchez}, {Torres}, {Casares},
  {Munoz-Darias}, {Armas Padilla}, \& {Yanes-Rizo}}]{2024arXiv240813310M}
{Mata Sanchez}, D., {Torres}, M.~A.~P., {Casares}, J., {et~al.} 2024, arXiv
  e-prints, arXiv:2408.13310, \dodoi{10.48550/arXiv.2408.13310}

\bibitem[{{McClintock} \& {Remillard}(2006)}]{2006csxs.book..157M}
{McClintock}, J.~E., \& {Remillard}, R.~A. 2006, in Compact stellar X-ray
  sources, ed. W.~H.~G. {Lewin} \& M.~{van der Klis}, Vol.~39, 157--213,
  \dodoi{10.48550/arXiv.astro-ph/0306213}

\bibitem[{{Mirabel} \& {Rodrigues}(2003)}]{2003Sci...300.1119M}
{Mirabel}, I.~F., \& {Rodrigues}, I. 2003, Science, 300, 1119,
  \dodoi{10.1126/science.1083451}

\bibitem[{{Moe} \& {Di Stefano}(2017)}]{2017ApJS..230...15M}
{Moe}, M., \& {Di Stefano}, R. 2017, \apjs, 230, 15,
  \dodoi{10.3847/1538-4365/aa6fb6}

\bibitem[{{M{\"u}ller} {et~al.}(2016){M{\"u}ller}, {Heger}, {Liptai}, \&
  {Cameron}}]{2016MNRAS.460..742M}
{M{\"u}ller}, B., {Heger}, A., {Liptai}, D., \& {Cameron}, J.~B. 2016, \mnras,
  460, 742, \dodoi{10.1093/mnras/stw1083}

\bibitem[{{Nagarajan} \& {El-Badry}(2024)}]{2024arXiv241116847N}
{Nagarajan}, P., \& {El-Badry}, K. 2024, arXiv e-prints, arXiv:2411.16847,
  \dodoi{10.48550/arXiv.2411.16847}

\bibitem[{{Naoz}(2016)}]{2016ARA&A..54..441N}
{Naoz}, S. 2016, \araa, 54, 441, \dodoi{10.1146/annurev-astro-081915-023315}

\bibitem[{{Naoz} \& {Fabrycky}(2014)}]{2014ApJ...793..137N}
{Naoz}, S., \& {Fabrycky}, D.~C. 2014, \apj, 793, 137,
  \dodoi{10.1088/0004-637X/793/2/137}

\bibitem[{{Naoz} {et~al.}(2016){Naoz}, {Fragos}, {Geller}, {Stephan}, \&
  {Rasio}}]{2016ApJ...822L..24N}
{Naoz}, S., {Fragos}, T., {Geller}, A., {Stephan}, A.~P., \& {Rasio}, F.~A.
  2016, \apjl, 822, L24, \dodoi{10.3847/2041-8205/822/2/L24}

\bibitem[{{Nugis} \& {Lamers}(2000)}]{2000A&A...360..227N}
{Nugis}, T., \& {Lamers}, H.~J.~G.~L.~M. 2000, \aap, 360, 227

\bibitem[{{{\"O}zel} {et~al.}(2010){{\"O}zel}, {Psaltis}, {Narayan}, \&
  {McClintock}}]{2010ApJ...725.1918O}
{{\"O}zel}, F., {Psaltis}, D., {Narayan}, R., \& {McClintock}, J.~E. 2010,
  \apj, 725, 1918, \dodoi{10.1088/0004-637X/725/2/1918}

\bibitem[{{Paczynski}(1976)}]{1976IAUS...73...75P}
{Paczynski}, B. 1976, in IAU Symposium, Vol.~73, Structure and Evolution of
  Close Binary Systems, ed. P.~{Eggleton}, S.~{Mitton}, \& J.~{Whelan}, 75

\bibitem[{{Pauli} {et~al.}(2022){Pauli}, {Langer}, {Aguilera-Dena}, {Wang}, \&
  {Marchant}}]{2022A&A...667A..58P}
{Pauli}, D., {Langer}, N., {Aguilera-Dena}, D.~R., {Wang}, C., \& {Marchant},
  P. 2022, \aap, 667, A58, \dodoi{10.1051/0004-6361/202243965}

\bibitem[{{Paxton} {et~al.}(2011){Paxton}, {Bildsten}, {Dotter}, {Herwig},
  {Lesaffre}, \& {Timmes}}]{2011ApJS..192....3P}
{Paxton}, B., {Bildsten}, L., {Dotter}, A., {et~al.} 2011, \apjs, 192, 3,
  \dodoi{10.1088/0067-0049/192/1/3}

\bibitem[{{Paxton} {et~al.}(2013){Paxton}, {Cantiello}, {Arras}, {Bildsten},
  {Brown}, {Dotter}, {Mankovich}, {Montgomery}, {Stello}, {Timmes}, \&
  {Townsend}}]{2013ApJS..208....4P}
{Paxton}, B., {Cantiello}, M., {Arras}, P., {et~al.} 2013, \apjs, 208, 4,
  \dodoi{10.1088/0067-0049/208/1/4}

\bibitem[{{Paxton} {et~al.}(2015){Paxton}, {Marchant}, {Schwab}, {Bauer},
  {Bildsten}, {Cantiello}, {Dessart}, {Farmer}, {Hu}, {Langer}, {Townsend},
  {Townsley}, \& {Timmes}}]{2015ApJS..220...15P}
{Paxton}, B., {Marchant}, P., {Schwab}, J., {et~al.} 2015, \apjs, 220, 15,
  \dodoi{10.1088/0067-0049/220/1/15}

\bibitem[{{Paxton} {et~al.}(2018){Paxton}, {Schwab}, {Bauer}, {Bildsten},
  {Blinnikov}, {Duffell}, {Farmer}, {Goldberg}, {Marchant}, {Sorokina},
  {Thoul}, {Townsend}, \& {Timmes}}]{2018ApJS..234...34P}
{Paxton}, B., {Schwab}, J., {Bauer}, E.~B., {et~al.} 2018, \apjs, 234, 34,
  \dodoi{10.3847/1538-4365/aaa5a8}

\bibitem[{{Paxton} {et~al.}(2019){Paxton}, {Smolec}, {Schwab}, {Gautschy},
  {Bildsten}, {Cantiello}, {Dotter}, {Farmer}, {Goldberg}, {Jermyn}, {Kanbur},
  {Marchant}, {Thoul}, {Townsend}, {Wolf}, {Zhang}, \&
  {Timmes}}]{2019ApJS..243...10P}
{Paxton}, B., {Smolec}, R., {Schwab}, J., {et~al.} 2019, \apjs, 243, 10,
  \dodoi{10.3847/1538-4365/ab2241}

\bibitem[{{Podsiadlowski} {et~al.}(1995){Podsiadlowski}, {Cannon}, \&
  {Rees}}]{1995MNRAS.274..485P}
{Podsiadlowski}, P., {Cannon}, R.~C., \& {Rees}, M.~J. 1995, \mnras, 274, 485,
  \dodoi{10.1093/mnras/274.2.485}

\bibitem[{{Qin} {et~al.}(2023){Qin}, {Hu}, {Meynet}, {Wang}, {Zhu}, {Song},
  {Shu}, \& {Wu}}]{2023A&A...671A..62Q}
{Qin}, Y., {Hu}, R.~C., {Meynet}, G., {et~al.} 2023, \aap, 671, A62,
  \dodoi{10.1051/0004-6361/202244712}

\bibitem[{{Rajamuthukumar} {et~al.}(2023){Rajamuthukumar}, {Hamers},
  {Neunteufel}, {Pakmor}, \& {de Mink}}]{2023ApJ...950....9R}
{Rajamuthukumar}, A.~S., {Hamers}, A.~S., {Neunteufel}, P., {Pakmor}, R., \&
  {de Mink}, S.~E. 2023, \apj, 950, 9, \dodoi{10.3847/1538-4357/acc86c}

\bibitem[{{Rantala} {et~al.}(2020){Rantala}, {Pihajoki}, {Mannerkoski},
  {Johansson}, \& {Naab}}]{2020MNRAS.492.4131R}
{Rantala}, A., {Pihajoki}, P., {Mannerkoski}, M., {Johansson}, P.~H., \&
  {Naab}, T. 2020, \mnras, 492, 4131, \dodoi{10.1093/mnras/staa084}

\bibitem[{{Remillard} \& {McClintock}(2006)}]{2006ARA&A..44...49R}
{Remillard}, R.~A., \& {McClintock}, J.~E. 2006, \araa, 44, 49,
  \dodoi{10.1146/annurev.astro.44.051905.092532}

\bibitem[{{Riley} {et~al.}(2022){Riley}, {Agrawal}, {Barrett}, {Boyett},
  {Broekgaarden}, {Chattopadhyay}, {Gaebel}, {Gittins}, {Hirai}, {Howitt},
  {Justham}, {Khandelwal}, {Kummer}, {Lau}, {Mandel}, {de Mink}, {Neijssel},
  {Riley}, {van Son}, {Stevenson}, {Vigna-G{\'o}mez}, {Vinciguerra}, {Wagg},
  {Willcox}, \& {Team Compas}}]{2022ApJS..258...34R}
{Riley}, J., {Agrawal}, P., {Barrett}, J.~W., {et~al.} 2022, \apjs, 258, 34,
  \dodoi{10.3847/1538-4365/ac416c}

\bibitem[{{R{\"o}pke} \& {De Marco}(2023)}]{2023LRCA....9....2R}
{R{\"o}pke}, F.~K., \& {De Marco}, O. 2023, Living Reviews in Computational
  Astrophysics, 9, 2, \dodoi{10.1007/s41115-023-00017-x}

\bibitem[{{Rostami-Shirazi} {et~al.}(2024){Rostami-Shirazi}, {Hasani Zonoozi},
  {Haghi}, \& {Rabiee}}]{2024arXiv241016276R}
{Rostami-Shirazi}, A., {Hasani Zonoozi}, A., {Haghi}, H., \& {Rabiee}, M. 2024,
  arXiv e-prints, arXiv:2410.16276, \dodoi{10.48550/arXiv.2410.16276}

\bibitem[{{Sabach} \& {Soker}(2015)}]{2015MNRAS.450.1716S}
{Sabach}, E., \& {Soker}, N. 2015, \mnras, 450, 1716,
  \dodoi{10.1093/mnras/stv717}

\bibitem[{{Salas} {et~al.}(2019){Salas}, {Naoz}, {Morris}, \&
  {Stephan}}]{2019MNRAS.487.3029S}
{Salas}, J.~M., {Naoz}, S., {Morris}, M.~R., \& {Stephan}, A.~P. 2019, \mnras,
  487, 3029, \dodoi{10.1093/mnras/stz1515}

\bibitem[{{Sana} {et~al.}(2013){Sana}, {de Koter}, {de Mink}, {Dunstall},
  {Evans}, {H{\'e}nault-Brunet}, {Ma{\'\i}z Apell{\'a}niz},
  {Ram{\'\i}rez-Agudelo}, {Taylor}, {Walborn}, {Clark}, {Crowther}, {Herrero},
  {Gieles}, {Langer}, {Lennon}, \& {Vink}}]{2013A&A...550A.107S}
{Sana}, H., {de Koter}, A., {de Mink}, S.~E., {et~al.} 2013, \aap, 550, A107,
  \dodoi{10.1051/0004-6361/201219621}

\bibitem[{{Schneider} {et~al.}(2021){Schneider}, {Podsiadlowski}, \&
  {M{\"u}ller}}]{2021A&A...645A...5S}
{Schneider}, F.~R.~N., {Podsiadlowski}, P., \& {M{\"u}ller}, B. 2021, \aap,
  645, A5, \dodoi{10.1051/0004-6361/202039219}

\bibitem[{{Shao} \& {Li}(2015)}]{2015ApJ...809...99S}
{Shao}, Y., \& {Li}, X.-D. 2015, \apj, 809, 99,
  \dodoi{10.1088/0004-637X/809/1/99}

\bibitem[{{Shao} \& {Li}(2020)}]{2020ApJ...898..143S}
---. 2020, \apj, 898, 143, \dodoi{10.3847/1538-4357/aba118}

\bibitem[{{Shao} \& {Li}(2021)}]{2021ApJ...920...81S}
---. 2021, \apj, 920, 81, \dodoi{10.3847/1538-4357/ac173e}

\bibitem[{{Shariat} {et~al.}(2024{\natexlab{a}}){Shariat}, {Naoz}, {El-Badry},
  {Akira Rocha}, {Kalogera}, {Stephan}, {Burdge}, \&
  {Angelo}}]{2024arXiv241115644S}
{Shariat}, C., {Naoz}, S., {El-Badry}, K., {et~al.} 2024{\natexlab{a}}, arXiv
  e-prints, arXiv:2411.15644, \dodoi{10.48550/arXiv.2411.15644}

\bibitem[{{Shariat} {et~al.}(2024{\natexlab{b}}){Shariat}, {Naoz}, {El-Badry},
  {Rodriguez}, {Hansen}, {Angelo}, \& {Stephan}}]{2024arXiv240706257S}
---. 2024{\natexlab{b}}, arXiv e-prints, arXiv:2407.06257,
  \dodoi{10.48550/arXiv.2407.06257}

\bibitem[{{Shariat} {et~al.}(2023){Shariat}, {Naoz}, {Hansen}, {Angelo},
  {Michaely}, \& {Stephan}}]{2023ApJ...955L..14S}
{Shariat}, C., {Naoz}, S., {Hansen}, B. M.~S., {et~al.} 2023, \apjl, 955, L14,
  \dodoi{10.3847/2041-8213/acf76b}

\bibitem[{{Shenar} {et~al.}(2022){Shenar}, {Sana}, {Mahy}, {El-Badry},
  {Marchant}, {Langer}, {Hawcroft}, {Fabry}, {Sen}, {Almeida}, {Abdul-Masih},
  {Bodensteiner}, {Crowther}, {Gieles}, {Gromadzki}, {H{\'e}nault-Brunet},
  {Herrero}, {de Koter}, {Iwanek}, {Koz{\l}owski}, {Lennon}, {Ma{\'\i}z
  Apell{\'a}niz}, {Mr{\'o}z}, {Moffat}, {Picco}, {Pietrukowicz}, {Poleski},
  {Rybicki}, {Schneider}, {Skowron}, {Skowron}, {Soszy{\'n}ski},
  {Szyma{\'n}ski}, {Toonen}, {Udalski}, {Ulaczyk}, {Vink}, \&
  {Wrona}}]{2022NatAs...6.1085S}
{Shenar}, T., {Sana}, H., {Mahy}, L., {et~al.} 2022, Nature Astronomy, 6, 1085,
  \dodoi{10.1038/s41550-022-01730-y}

\bibitem[{{Soker}(2021)}]{2021MNRAS.504.5967S}
{Soker}, N. 2021, \mnras, 504, 5967, \dodoi{10.1093/mnras/stab1275}

\bibitem[{{Stegmann} {et~al.}(2022){Stegmann}, {Antonini}, \&
  {Moe}}]{2022MNRAS.516.1406S}
{Stegmann}, J., {Antonini}, F., \& {Moe}, M. 2022, \mnras, 516, 1406,
  \dodoi{10.1093/mnras/stac2192}

\bibitem[{{Sukhbold} {et~al.}(2016){Sukhbold}, {Ertl}, {Woosley}, {Brown}, \&
  {Janka}}]{2016ApJ...821...38S}
{Sukhbold}, T., {Ertl}, T., {Woosley}, S.~E., {Brown}, J.~M., \& {Janka}, H.~T.
  2016, \apj, 821, 38, \dodoi{10.3847/0004-637X/821/1/38}

\bibitem[{{Tanikawa} {et~al.}(2023){Tanikawa}, {Hattori}, {Kawanaka},
  {Kinugawa}, {Shikauchi}, \& {Tsuna}}]{2023ApJ...946...79T}
{Tanikawa}, A., {Hattori}, K., {Kawanaka}, N., {et~al.} 2023, \apj, 946, 79,
  \dodoi{10.3847/1538-4357/acbf36}

\bibitem[{{Toonen} {et~al.}(2016){Toonen}, {Hamers}, \& {Portegies
  Zwart}}]{2016ComAC...3....6T}
{Toonen}, S., {Hamers}, A., \& {Portegies Zwart}, S. 2016, Computational
  Astrophysics and Cosmology, 3, 6, \dodoi{10.1186/s40668-016-0019-0}

\bibitem[{{Toonen} {et~al.}(2018{\natexlab{a}}){Toonen}, {Perets}, \&
  {Hamers}}]{2018A&A...610A..22T}
{Toonen}, S., {Perets}, H.~B., \& {Hamers}, A.~S. 2018{\natexlab{a}}, \aap,
  610, A22, \dodoi{10.1051/0004-6361/201731874}

\bibitem[{{Toonen} {et~al.}(2018{\natexlab{b}}){Toonen}, {Perets}, {Igoshev},
  {Michaely}, \& {Zenati}}]{2018A&A...619A..53T}
{Toonen}, S., {Perets}, H.~B., {Igoshev}, A.~P., {Michaely}, E., \& {Zenati},
  Y. 2018{\natexlab{b}}, \aap, 619, A53, \dodoi{10.1051/0004-6361/201833164}

\bibitem[{{Toonen} {et~al.}(2020){Toonen}, {Portegies Zwart}, {Hamers}, \&
  {Bandopadhyay}}]{2020A&A...640A..16T}
{Toonen}, S., {Portegies Zwart}, S., {Hamers}, A.~S., \& {Bandopadhyay}, D.
  2020, \aap, 640, A16, \dodoi{10.1051/0004-6361/201936835}

\bibitem[{{Tout} {et~al.}(1997){Tout}, {Aarseth}, {Pols}, \&
  {Eggleton}}]{1997MNRAS.291..732T}
{Tout}, C.~A., {Aarseth}, S.~J., {Pols}, O.~R., \& {Eggleton}, P.~P. 1997,
  \mnras, 291, 732, \dodoi{10.1093/mnras/291.4.732}

\bibitem[{{van den Heuvel}(1976)}]{1976IAUS...73...35V}
{van den Heuvel}, E.~P.~J. 1976, in IAU Symposium, Vol.~73, Structure and
  Evolution of Close Binary Systems, ed. P.~{Eggleton}, S.~{Mitton}, \&
  J.~{Whelan}, 35

\bibitem[{{Vetter} {et~al.}(2024){Vetter}, {R{\"o}pke}, {Schneider}, {Pakmor},
  {Ohlmann}, {Lau}, \& {Andrassy}}]{2024A&A...691A.244V}
{Vetter}, M., {R{\"o}pke}, F.~K., {Schneider}, F. R.~N., {et~al.} 2024, \aap,
  691, A244, \dodoi{10.1051/0004-6361/202451579}

\bibitem[{{Vigna-G{\'o}mez} {et~al.}(2024){Vigna-G{\'o}mez}, {Willcox},
  {Tamborra}, {Mandel}, {Renzo}, {Wagg}, {Janka}, {Kresse}, {Bodensteiner},
  {Shenar}, \& {Tauris}}]{2024PhRvL.132s1403V}
{Vigna-G{\'o}mez}, A., {Willcox}, R., {Tamborra}, I., {et~al.} 2024, \prl, 132,
  191403, \dodoi{10.1103/PhysRevLett.132.191403}

\bibitem[{{Vink} {et~al.}(2001){Vink}, {de Koter}, \&
  {Lamers}}]{2001A&A...369..574V}
{Vink}, J.~S., {de Koter}, A., \& {Lamers}, H.~J.~G.~L.~M. 2001, \aap, 369,
  574, \dodoi{10.1051/0004-6361:20010127}

\bibitem[{{von Zeipel}(1910)}]{1910AN....183..345V}
{von Zeipel}, H. 1910, Astronomische Nachrichten, 183, 345,
  \dodoi{10.1002/asna.19091832202}

\bibitem[{{Walk} {et~al.}(2020){Walk}, {Tamborra}, {Janka}, {Summa}, \&
  {Kresse}}]{2020PhRvD.101l3013W}
{Walk}, L., {Tamborra}, I., {Janka}, H.-T., {Summa}, A., \& {Kresse}, D. 2020,
  \prd, 101, 123013, \dodoi{10.1103/PhysRevD.101.123013}

\bibitem[{{Wang} {et~al.}(2024){Wang}, {Zhao}, {Feng}, {Ge}, {Shao}, {Cui},
  {Gao}, {Zhang}, {Wang}, {Li}, {Bai}, {Yuan}, {Huang}, {Yuan}, {Zhang}, {Yi},
  {Xiang}, {Li}, {Li}, {Zhang}, {Zhang}, {Han}, {Fan}, {Li}, {Chen}, {Liu},
  {Meng}, {Liu}, {Zhang}, {Gu}, \& {Liu}}]{2024NatAs.tmp..215W}
{Wang}, S., {Zhao}, X., {Feng}, F., {et~al.} 2024, Nature Astronomy,
  \dodoi{10.1038/s41550-024-02359-9}

\bibitem[{{Webbink}(1984)}]{1984ApJ...277..355W}
{Webbink}, R.~F. 1984, \apj, 277, 355, \dodoi{10.1086/161701}

\bibitem[{{Xu} \& {Li}(2010)}]{2010ApJ...716..114X}
{Xu}, X.-J., \& {Li}, X.-D. 2010, \apj, 716, 114,
  \dodoi{10.1088/0004-637X/716/1/114}

\bibitem[{{Zorotovic} {et~al.}(2010){Zorotovic}, {Schreiber}, {G{\"a}nsicke},
  \& {Nebot G{\'o}mez-Mor{\'a}n}}]{2010A&A...520A..86Z}
{Zorotovic}, M., {Schreiber}, M.~R., {G{\"a}nsicke}, B.~T., \& {Nebot
  G{\'o}mez-Mor{\'a}n}, A. 2010, \aap, 520, A86,
  \dodoi{10.1051/0004-6361/200913658}

\end{thebibliography}
\bibliographystyle{./aasjournal}
\end{document}